\documentclass[journal]{new-aiaa}
\usepackage[utf8]{inputenc}
\usepackage{textcomp}

\usepackage{multirow}
\usepackage[table]{xcolor}
\usepackage{comment} 
\usepackage{authblk}
\usepackage{algorithmicx}
\usepackage[ruled]{algorithm2e}% http://ctan.org/pkg/algorithmicx
\usepackage{url}
\usepackage{caption}
\usepackage{subcaption}

\allowdisplaybreaks[4]
\newcommand\mcedit[1]{\textcolor{blue}{#1}}
\newcommand\mdsedit[1]{\textcolor{blue}{#1}}

\usepackage{graphicx}
\usepackage{amsmath}
\usepackage[version=4]{mhchem}
\usepackage{siunitx}
\usepackage{longtable,tabularx}
\title{Data-driven robust MPC of tiltwing VTOL aircraft}

\author{Martin Doff-Sotta \footnote{Postdoctoral Researcher, Department of Engineering Science, martin.doff-sotta@eng.ox.ac.uk}, Mark Cannon\footnote{Associate Professor, Department of Engineering Science, mark.cannon@eng.ox.ac.uk} and Marko Bacic\footnote{Engineering Fellow - Control Systems, Civil Aerospace, Rolls-Royce, Plc.University Research Lecturer, University of Oxford, marko.bacic@rolls-royce.com}}
\affil{University of Oxford, Parks Road, Oxford OX1 3PJ, United Kingdom}

\begin{document}

\maketitle

\begin{abstract}
This paper investigates robust tube-based Model Predictive Control (MPC) of a tiltwing Vertical Take-Off and Landing (VTOL) aircraft subject to wind disturbances and model uncertainty. Our approach is based on a Difference of Convex (DC) function decomposition of the dynamics to develop a computationally tractable optimisation with robust tubes for the system trajectories. We consider a case study of a VTOL aircraft subject to wind gusts and whose aerodynamics is defined from data.
\end{abstract}

\begin{comment}

\section*{Nomenclature}

\noindent(Nomenclature entries should have the units identified)

{\renewcommand\arraystretch{1.0}
\noindent\begin{longtable*}{@{}l @{\quad=\quad} l@{}}
$A$  & amplitude of oscillation \\
$a$ &    cylinder diameter \\
$C_p$& pressure coefficient \\
$Cx$ & force coefficient in the \textit{x} direction \\
$Cy$ & force coefficient in the \textit{y} direction \\
c   & chord \\
d$t$ & time step \\
$Fx$ & $X$ component of the resultant pressure force acting on the vehicle \\
$Fy$ & $Y$ component of the resultant pressure force acting on the vehicle \\
$f, g$   & generic functions \\
$h$  & height \\
$i$  & time index during navigation \\
$j$  & waypoint index \\
$K$  & trailing-edge (TE) nondimensional angular deflection rate\\
$\Theta$ & boundary-layer momentum thickness\\
$\rho$ & density\\
\multicolumn{2}{@{}l}{Subscripts}\\
cg & center of gravity\\
$G$ & generator body\\
iso	& waypoint index
\end{longtable*}}
\end{comment}

\section{Introduction}

Urban Air Mobility (UAM) has a potential to transform transportation of people and goods in congested cities \cite{easa} while potentially reducing ground traffic. A key enabler of this technology is a class of zero carbon emission eVTOL aircraft concepts \cite{kadhiresan2019conceptual}, many based on  tiltrotor, tiltduct or tiltwing vehicle configurations powered by batteries or by making use of zero carbon fuel like hydrogen \cite{SAIAS202232655}. However, the lower energy density of energy storage options relative to liquid carbon based fuels restrict the operational range, hover time and cruise speeds. Additionally such concepts require transition   between vertical and horizontal flight further complicating operational scenarios. Consequently limited energy and transition envelope constraints impede the throughput of flight operations by limiting the number of flights, timely allocation of landing slots, duration of holding patterns in hover, and separation between vehicles. To maximise throughput it is clear that both the energy spent in a non-wing-borne flight phase and the time and space needed to transition between thrust-borne and wing-borne flight should be minimised.

This paper proposes a robust control methodology for transitions of  VTOL aircraft between wing-borne and thrust-borne flight phases in the presence of wind gusts, model uncertainty and state constraints.  We explore a Model Predictive Control framework due to its potential to outperform classical control architectures by optimising future control sequences with respect to a specified objective (usually minimum time or minimum energy transition) with explicit constraint handling using a model of the vehicle. This methodology can also provide guarantees of robustness to model uncertainty and external disturbances \cite{kouvaritakis2016model}. Although we develop here MPC policies for a generic class of tiltwing aircraft, the ideas presented in this paper are equally applicable to tilt rotors and tilt ducts.
%% These vehicles capable of vertical take-off and landing in restricted spaces (Fig. \ref{fig:eVTOL})are particularly appealing in UAM scenarios but the complex dynamics and stability issues inherent to the transition from thrust- to wing- borne flight are a barrier to their widespread adoption in a safety-critical context. 

%\begin{figure}[ht]
%    \centering
%    \includegraphics[width=0.6\textwidth]{img/eVTOL.jpeg} %{<left> <lower> <right> <upper>}
%    \caption{Prototype eVTOL aircraft for passenger transport developed by Rolls-Royce as part of its UAM program.}
%    \label{fig:eVTOL}
%\end{figure}
%We consider an emerging application from Urban Air Mobility (UAM) to demonstrate the viability of the method in solving realistic large-scale problems. In particular, this research investigates computing optimal trajectories for the transition of a tiltwing VTOL aircraft capable of vertical take-off and landing subject to constraints. 

%

Various approaches have been proposed to address this problem. In \cite{me}, a cascaded PID control architecture was proposed for the transition of a prototype tiltrotor VTOL aircraft. The transition is achieved through smooth scheduling functions of the forward velocity or tilt angle, and the simulation results are supported by flight test results. Modeling, control and flight testing for the transition of a tiltwing aircraft are achieved in \cite{Siegwart2019}, extending the P-PID structure from the PX4 open-source software with feedback linearisation, gain scheduling and model-based control allocation. A gain-scheduled LQR control architecture is presented in \cite{EUCASS} for the transition of a tandem tiltwing aircraft in the presence of moderate wind gusts. Instabilities were observed for wind gusts of intensity larger than $5 \si{m/s}$ during transitions, which limits the viability of the approach in the presence of wind in realistic conditions.

Recent advances in robust tube-based MPC allow robust control of nonlinear systems whose dynamics can be represented as a difference of convex functions \cite{DC-TMPC}. The main idea is to successively linearise the dynamics around predicted trajectories and treat linearisation errors as bounded disturbances. Because the linearised functions are convex, so are their linearisation errors, and since these errors are maximised at the boundary of the domain on which they are evaluated, they can therefore be bounded tightly. The trajectories of model states are thus bounded tightly by a sequence of sets (known as a tube \cite{kouvaritakis2016model}) defined by convex inequalities. Although very efficient, the scope of applicability is initially limited to systems with convex dynamics. However, we show that if the dynamics are sufficiently regular, techniques from difference of convex (DC) decomposition of polynomials \cite{ahmadi} can be used to represent the VTOL nonlinear dynamics as a difference of convex functions, which allows the powerful approach in \cite{DC-TMPC} to be used. The dynamics need not be derived using first principles modelling; we employ a mixture of data-based and physical models. The proposed approach is the culmination of our work~\cite{ACC, CCTA} on convex trajectory optimisation of VTOL aircraft.

%In \cite{UMich19}, the trajectory optimisation for the transition of the Airbus A$^3$ Vahana is formulated as a constrained nonlinear optimisation problem. The project addresses important modeling aspects such as the beyond stall drag and lift curves and wing flow augmentation due to wing-propeller interaction.  In \cite{ACC}, the transition for a tiltwing VTOL aircraft was computed using convex optimisation by introducing a small angle approximation. This provides a computationally efficient optimisation that could potentially be leveraged online, e.g. for collision avoidance or MPC. The obvious limitation of the approach is the assumption of small angles of attack, which restricts considerably the type of achievable manoeuvres.  

The specific contributions of this research over earlier work are as follows: i) we propose a computationally tractable, optimal, robust control architecture for VTOL aircraft subject to additive disturbance and model uncertainty; ii) we combine DC decomposition with robust tube-based MPC and demonstrate the applicability and generalisability of the procedure in \cite{DC-TMPC}; iii) we show that our technique also applies when parts of the model are defined from data. 

This paper is organised as follows. We start by developing a mathematical model of a tiltwing VTOL aircraft subject to wind disturbance in Section \ref{sec:modeling}. In Section \ref{sec:TMPC}, we formulate the MPC optimisation problem and leverage a DC decomposition of the nonlinear dynamics to construct robust tubes for the state trajectories. Section \ref{sec:results} discusses simulation results obtained for a case study based on the Airbus A$^3$ Vahana. Section~\ref{sec:conclusion} presents conclusions.

\section{Modelling}
\subsection{Assumptions}
We assume a flat earth model and consider trajectory optimisation in the longitudinal plane alone. Tiltwing aircraft are considered (see Figure~\ref{fig:diagram}), with one or more wing surfaces carrying thrust effectors that can be rotated by an actuator through 90 degrees as the aircraft transitions between wing-borne and thrust-borne flight. We further assume classical inner/outer loop flight control laws for stabilisation of aircraft attitude with time-scale separation, so that the closed loop pitch dynamics are much faster than those of the desired flight path and tiltwing actuation. Consequently the pitch angle  $\theta$ (defined here as the angle of the fuselage axis from horizontal earth plane) is assumed to be maintained at all times by the attitude control loop at a constant reference angle $\theta^r = 0$. 
\label{sec:modeling}
\subsection{Equations of motion}
\begin{figure}[ht]
    \centering
    \includegraphics[width=0.7\textwidth, trim={1.5cm 3.5cm 1.5cm 3cm}, clip]{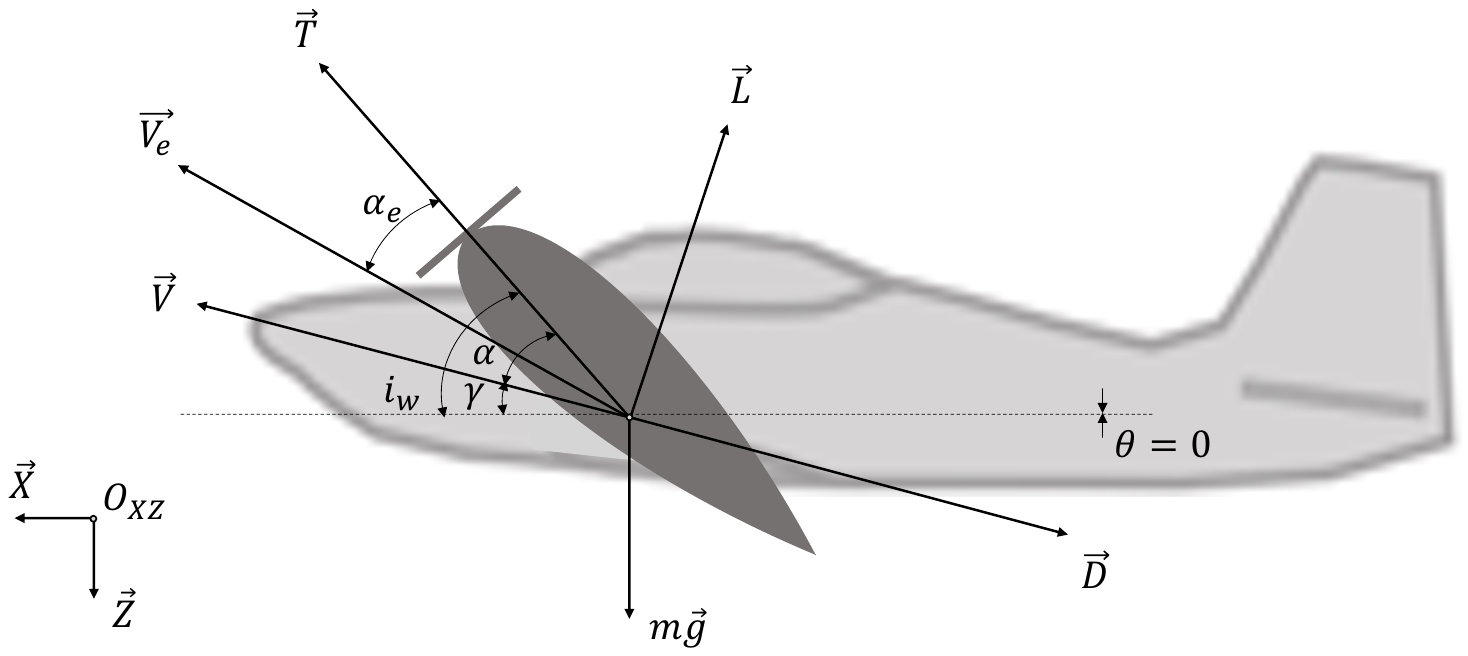} %{<left> <lower> <right> <upper>}
    \caption{Force and velocity definitions for a VTOL aircraft.}
    \label{fig:diagram}
\end{figure}

Consider a longitudinal point-mass model of a tiltwing VTOL aircraft equipped with propellers (Figure~\ref{fig:diagram}) subject to a wind gust disturbance. The Equations Of Motion (EOM) with respect to inertial frame $O_{XZ}$ are given by 
\begin{gather}
\label{eq:pos}
\dot{X} = V_x,
\quad
\dot{Z} = V_z,
\\
\label{eq:eom1}
 m \dot{V}_x = \underbrace{T \cos{(\alpha + \gamma)} - D \cos{\gamma} - L \sin {\gamma}}_{f_1} + W_x, 
\\
m  \dot{V}_z = \underbrace{-T \sin{(\alpha + \gamma)} + D \sin {\gamma} - L \cos {\gamma} + mg}_{f_2} + W_z, 
\label{eq:eom2}
\end{gather}
where $T$ is the thrust magnitude, $L$, $D$ are the lift and drag forces, $V_x, V_z$ the components of the aircraft velocity in the inertial frame $V = \sqrt{V_x^2 + V_z^2}$, $\alpha$ is the angle of attack, $\gamma = -\arcsin(V_z/V)$ is the flight path angle  (defined as the angle of the velocity vector from horizontal), and $(X, Z)$ the position in inertial frame.  Wind gusts are modelled by additive bounded disturbances $W_x$ and $W_z$, which are assumed to lie at all times within known bounds.

\begin{figure}[ht]
    \centering
    \includegraphics[width=0.6\textwidth]{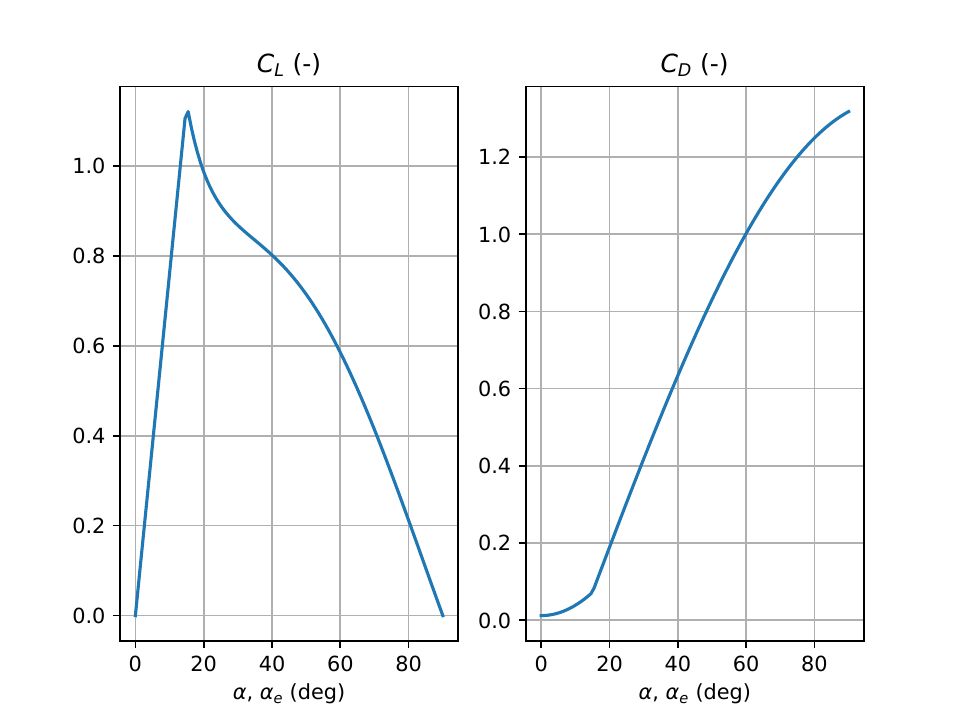} %{<left> <lower> <right> <upper>}
    \caption{Lift and drag coefficients as a function of angle of attack.}
    \label{fig:coef}
\end{figure}

The dynamics of the rotating wing are given by
\begin{equation}
J_w \dot{\zeta} = M, \quad \dot{i}_w = \zeta,
\label{eq:tiltwing_dyna}
\end{equation}
where $J_w$ is the rotational inertia of the wing (about the $y$-axis), $M$ is the total torque delivered by the tilting actuators, $\zeta$ is the tiltwing rate and $i_w$ is the tiltwing angle. The angles $i_w$, $\theta$, $\alpha$ and $\gamma$ are related by (see Figure \ref{fig:diagram})
\begin{equation}
\label{eq:cstr1}
i_w + \theta = \alpha + \gamma. 
\end{equation}
From momentum theory, the propeller generates an induced speed $v_i$ that is implicitly defined by
\[
\rho A n (V \cos{\alpha} + v_i )( k_w v_i) -T = 0,
\]
%
\begin{comment}
\[
v_i = -\frac{V \cos{\alpha}}{2} + \sqrt{\frac{V^2 \cos^2{\alpha}}{4} + \frac{T}{2\rho A n}},
\]
\end{comment}
%
where $\rho$ is the air density, $A$ the rotor disk area, $n$ the number of propellers, and $k_w \approx 2$~\cite{ACC}. The effective (blown) velocity $V_e$ and effective (blown) angle of attack $\alpha_e$ seen by the wing due to the effect of the propeller wake on the wing are given by
\begin{gather}
V_e \cos{\alpha_e} = V \cos{\alpha} + k_w v_i ,
\\
V_e \sin{\alpha_e} = V \sin{\alpha},
\\
V_e^2 = V^2 + \frac{2T}{\rho A n}. 
\end{gather}
The total lift and drag are modeled as the weighted sum of the blown and unblown terms as follows
\begin{gather}
\label{eq:L}
L = \tfrac{1}{2}  \lambda \rho S C_L(\alpha_e) V_e^2 + \tfrac{1}{2} (1-\lambda)  \rho S C_L(\alpha) V^2, 
\\
\label{eq:D}
D =\tfrac{1}{2} \lambda\rho S C_D(\alpha_e) V_e^2 + \tfrac{1}{2}  (1-\lambda) \rho S C_D(\alpha) V^2, 
\end{gather}
where $S$ is the wing area,  $\lambda$ is a weighting term representing the portion of the wing in the wake, and $C_L$ and $C_D$ are lift and drag coefficients. To illustrate the approach we define $C_L$ and $C_D$ using data derived from the Tangler-Ostowari post-stall model~\cite{UMich19} (see Figure \ref{fig:coef}), which determines lift and drag forces over a wide range of $\alpha$ and $\alpha_e$ values as the wing operates at high angles of attack during transitions.

\subsection{Constraints}
For the problem considered in this paper we assume that the gust disturbances $W_x, W_z$ in the forward and vertical directions are bounded: $W_i\in[\underline{W}_i,\overline{W}_i]$ for $i=x,z$.
Since wing tilting actuators have finite torque capacity we also assume bounded wing acceleration/deceleration rates given by $\underline{M}$ and $\overline{M}$. To ensure sensible trajectories for passenger comfort and g-loads, we further introduce constraint limits on horizontal ($\dot{V}_x$) and vertical ($\dot{V}_z$) accelerations. Finally, constraints on thrust range, tilt angle range and absolute velocities can all be expressed in the compact form as input and state constraints \cite{ACC},
\begin{gather}
\label{eq:cstr1bis}
\underline{V}_x \leq V_x \leq \overline{V}_x, \quad \underline{V}_z \leq V_z \leq \overline{V}_z, \quad 0 \leq T \leq \overline{T}, \quad \underline{i}_w \leq i_w \leq \overline{i}_w,
\\
\label{eq:cstr2}
V_x(t_0) = V_{x, 0}, \quad V_z(t_0) = V_{z, 0}, \quad i_w(t_0) = i_{0}, \\
\underline{W}_x \leq W_x \leq \overline{W}_x, \quad \underline{W}_z \leq W_z \leq \overline{W}_z, \\
\label{eq:cstr3bis}
\underline{a} \leq \dot{V}_x \leq \overline{a},  \quad \underline{a} \leq \dot{V}_z \leq \overline{a},  \\
\label{eq:second_order}
{\underline{M}}/{J_w} \leq \ddot{i}_w \leq {\overline{M}}/{J_w}.
\end{gather}

\section{Robust MPC formulation}
\label{sec:TMPC}

We now introduce a robust predictive control law for the system presented in Section \ref{sec:modeling}. Assuming we are only concerned with velocity control, the states $(X, Z)$ can be computed \textit{a posteriori} via \eqref{eq:pos} and thus eliminated from the analysis. Moreover, combining equations \eqref{eq:eom1}-\eqref{eq:D}, we can further eliminate the flight path angle, angle of attack, and  tiltwing rate from the formulation and express the dynamics with only two states $V_x,  V_z$ and two inputs $i_w, T$ as follows
\begin{gather}
\label{eq:eom_NDC1}
m\dot{V_x} = f_1 (V_x, V_z, i_w,  T) + W_x, \\
m\dot{V_z} = f_2 (V_x, V_z, i_w,  T) + W_z, 
\label{eq:eom_NDC2}
\end{gather}
where $i_w$ is now an input constrained by its second order derivative (see equation \ref{eq:second_order}). 

To setup the MPC problem we use the trajectory constraints given by (\ref{eq:cstr1bis})-(\ref{eq:second_order})
together with the terminal set for the velocities\cite{DC-TMPC}
\begin{equation}
\Biggl\| \Biggl[\begin{matrix} {V}_{x}(t_f) - V_{x}^r(t_f) \\ {V}_{z}(t_f) - V_{z}^r(t_f) \end{matrix}\Biggr] \Biggr\|_{\hat{Q}}^2 \leq  \hat{\gamma} ,
\label{eq:cstr3}
\end{equation}
where the notation $\cdot^r$ was used to denote a reference to be tracked, $t_f$ is a fixed terminal time, $\hat{\gamma}$ and $\hat{Q} \succ 0$ are respectively a terminal set bound and penalty matrix that can be computed following Appendix of \cite{DC-TMPC}. Note that equation \eqref{eq:cstr3} enforces the conditions: $|V_x(t_f) - V_{x}^r(t_f)| \leq \delta^{V}$, $|V_z(t_f) - V_{z}^r(t_f)| \leq \delta^{V}$, $|i_w(t_f) - i_w^r(t_f)| \leq \delta^{i_w}, \quad |T(t_f) - T^r(t_f)| \leq \delta^{T}$, where $\delta^{V}, \delta^{i_w}, \delta^{T}$, are terminal sets bounds. 

The control objective is to achieve tracking of a reference trajectory $V_x^r, V_z^r, i_w^r, T^r$ while rejecting the wind disturbances $W_x, W_z$ acting on the system. To do so, at each time step, we could compute a receding horizon control law that minimises the worst case quadratic objective defined for $Q_x \succ 0$, $Q_u\succ 0$ by 
\begin{equation}
\begin{aligned}
    \label{eq:obj}
    J(u) = 
\max_{\substack{W_x\in[ \underline{W}_x, \overline{W}_x] \\ 
W_z\in [\underline{W}_z, \overline{W}_z]}}
\Biggl\{ 
&
\Biggl\| \Biggl[\begin{matrix} V_x(i_w(t_f), T(t_f), W_x(t_f), W_z(t_f)) - V_x^r(t_f) 
\\ V_z(i_w(t_f), T(t_f), W_x(t_f), W_z(t_f)) - V_z^r(t_f) \end{matrix}\Biggr] \Biggr\|_{\hat{Q}}^2 
\\
&
+ \int_{t_0}^{t_f} \Biggl\| \Biggl[ \begin{matrix} V_x(i_w(t), T(t), W_x(t), W_z(t)) - V_x^r(t) \\ V_z(i_w(t), T(t), W_x(t), W_z(t)) - V_z^r(t) \end{matrix}\Biggr] \Biggr\|_{Q_x}^2 
+ \ \ \Biggl\| \Biggl[ \begin{matrix} i_w(t) - i_w^r(t) \\ T(t) - T^r(t)\end{matrix} \Biggr] \Biggr\|_{Q_u}^2 \, \mathrm{d}t \Biggr\}
\end{aligned}
\end{equation}
subject to \eqref{eq:cstr1bis}-\eqref{eq:cstr3}. Note that the states are uncertain and that the state trajectories in the objective are computed as the worst case realisation of the time-varying gust disturbances $W_x, W_z$. We would then apply the first element of the obtained optimal control sequence, update the current state and input and repeat the process at each time step. Since the model includes the nonlinear functions $f_1$ and $f_2$ and that the state is uncertain, computing the control law would require solving a min-max Nonlinear Program (NLP) which is intractable in practice. In what follows, we introduce a DC decomposition of these nonlinear functions in order to obtain a computationally efficient implementation of the robust MPC problem.

\subsection{DC decomposition}
\label{sub:DC}

Motivated by the fact that convex functions can be bounded tightly by convex and linear inequalities (as in \cite{DC-TMPC}), we seek DC decompositions of $f_1, f_2$: $f_1 = g_1 -h_1$ and $f_2 = g_2 -h_2$, where $g_1, h_1, g_2, h_2$ are convex. A DC decomposition always exists if $f_1, f_2 \in \mathcal{C}^2$ \cite{hartman} and can be precomputed offline. A similar procedure was first presented in \cite{CCTA} and follows an idea from \cite{ahmadi} on DC decomposition of nonconvex polynomials using algebraic techniques. In what follows we detail the procedure for the DC decomposition of a general function $f$ (and hence its applicability to $f_1$ and $f_2$): 
\subsubsection{Fit polynomial to data} Assume that the nonlinear model\footnote{Note that it does not need to be a mathematical function but can be defined from data. In the present case, $f$ is partly defined from data through the lift and drag coefficients in Figure \ref{fig:coef}.} $f$ can be approximated arbitrarily closely by a polynomial of degree $2d$ in Gram form such that $f \approx y(x)^\top P y(x)$, where $P = P^\top$ is the Gram matrix and $y = [1, x_1, x_2, \dots, x_n, x_1 x_2, \dots, x_n^{d}]^\top$ is a vector of monomials of degree up to $d$ ($y$ has size $C^{|x|}_{d + |x|}$). Generate $N_s$ samples $F_{s} = f(x_s), \forall s \in [1, ..., N_s]$  of the nonlinear model and solve the following least squares problem: 
\begin{equation*}
\mathcal{LS} : \quad \min_{\substack{P}} \quad  \sum_{s=0}^{N_s} \lVert F_{s} - y(x_s)^\top P y(x_s) \rVert^2_2,  \quad \text{ s.t.} \quad P = P ^\top. 
\end{equation*}

\subsubsection{Compute the Hessians of the decomposition} Let $g(x) \approx y(x)^\top Q y(x)$ and $h(x) \approx y(x)^\top R y(x)$ be convex polynomials such that their Hessians $d^2g(x)/d x^2 = y(x)^\top H_{g} y(x)$ and $d^2h(x)/d x^2 = y(x)^\top H_{h} y(x)$ are Positive Semi-Definite (PSD). Finding $Q, R$ such that $P = Q - R$ and $h(x)$, $g(x)$ are convex reduces to solving the following Semi Definite Program (SDP)
\begin{equation}
\begin{aligned}
 & \mathcal{SDP} : \max_{\substack{Q, \sigma}} & & \sigma \quad  \text{ s.t.} & & H_g(Q) \succeq \sigma I  ,   \quad H_h(Q) \succeq \sigma I , 
\nonumber
\end{aligned}
\end{equation}
with $\forall i, j$
\begin{gather*}
    [H_{g}]_{ij} = {D}_{j, i}^\top Q + Q {D}_{i, j} + {D}_i^\top Q D_j  + {D}_j^\top Q D_i, \\
[H_{h}]_{ij} = {D}_{j, i}^\top (Q - P)+ (Q - P) {D}_{i, j} + {D}_i^\top (Q - P) D_j  + {D}_j^\top (Q - P) D_i, 
\end{gather*}
 where $I$ is the identity matrix of compatible dimensions, $D_i$ is a matrix of coefficients such that $dy/d {x_i} = D_i y $ and $D_{i, j} = D_i D_j$.

\subsection{Successive convex programming tube MPC for DC systems}
\label{sub:DC-TMPC}

The nonlinear dynamics in \eqref{eq:eom_NDC1}-\eqref{eq:eom_NDC2} can now be expressed in a DC form as follows
\begin{gather}
\label{eq:eom_DC1}
m\dot{V_x} = g_1 (V_x, V_z, i_w,  T) - h_1 (V_x, V_z, i_w,  T) + W_x, \\
m\dot{V_z} = g_2 (V_x, V_z, i_w,  T) - h_2 (V_x, V_z, i_w,  T) + W_z, 
\label{eq:eom_DC2}
\end{gather}
and the DC-TMPC algorithm presented in \cite{DC-TMPC} can be applied to the system.

In what follows we will exploit the convexity properties of the functions $g_1, h_1, g_2, h_2$ in \eqref{eq:eom_DC1}-\eqref{eq:eom_DC2} to approximate the dynamics by a set of convex inequalities with tight bounds on the state trajectories. To do so, we linearise the dynamics successively around feasible guessed trajectories and treat the linearisation error as a bounded disturbance  \cite{DC-TMPC}. We use the fact that the linearisation error of a convex (resp. concave) function is also convex (resp. concave) and can thus be bounded tightly. %since its maximum (resp. minimum) occurs at the boundary of the set on which the function is constrained. 
This allows us to construct a robust optimisation using the tube-based MPC framework \cite{kouvaritakis2016model}, and to obtain solutions that are robust to the model error introduced by the linearisation (i.e. model uncertainty) and to wind gusts (i.e. exogeneous additive disturbances). 

The  DC-TMPC framework is based on the following ingredients:

\subsubsection{Parameterisation of the control input} We start by assuming the following two-degree of freedom parameterisation of the control inputs as follows \cite{kouvaritakis2016model}
\begin{gather*}
    i_w = \mu + K_{i_w} (V_x - V_x^\circ) +  K_{i_w}' (V_z - V_z^\circ), \\ T = \tau +K_{T} (V_x - V_x^\circ) +  K_{T}' (V_z - V_z^\circ),
\end{gather*}
where $V_x^\circ, V_z^\circ$ are guess trajectories for the states, $\mu$, $\tau$  are feedforward terms (solution of the MPC optimisation stated in Section \ref{sub:discrete}) and $K_{i_w}, K_{i_w}', K_{T}, K_{T}'$ are feedback gains to be computed e.g. by solving a LQR problem for the linearised nominal ($W_x, W_z = 0$) system \eqref{eq:eom_DC1}-\eqref{eq:eom_DC2}. Note that $g_1, h_1, g_2, h_2$ defined in \eqref{eq:eom_DC1}-\eqref{eq:eom_DC2} are now functions of $V_x, V_z, \mu,  \tau$. 

\subsubsection{Successive linearisations} We assume the existence\footnote{We can obtain feasible initial trajectories by simulating the nominal aircraft dynamics with a prior-determined control law, such as PID, and checking \textit{a posteriori} that other constraints are satisfied. An alternative method is to solve an initial feasibility problem as discussed in~\cite{DC-TMPC}.} of a set of feasible guess trajectories $V_x^\circ, V_z^\circ, \mu^\circ, \tau^\circ$ for \eqref{eq:eom_DC1}-\eqref{eq:eom_DC2} and successively linearise the dynamics around the guessed trajectories. The Taylor series expansion of the nonlinear dynamics is given by $\forall i \in \{1, 2\}$
\begin{gather*}
g_i = \lfloor g_i \rfloor_{(V_x^\circ, V_z^\circ, \mu^\circ, \tau^\circ)} +  \lceil g_i \rceil_{(V_x^\circ, V_z^\circ, \mu^\circ, \tau^\circ)},\\
h_i = \lfloor h_i \rfloor_{(V_x^\circ, V_z^\circ, \mu^\circ, \tau^\circ)} +  \lceil h_i \rceil_{(V_x^\circ, V_z^\circ, \mu^\circ, \tau^\circ)},  
\end{gather*}
where the notation $\lfloor f \rfloor_{x^\circ} = f(x^\circ) + \nabla f^{\top}(x^\circ ) (x - x^\circ)$ stands for the Jacobian linear approximation of $f$ around $x^\circ$ and $ \lceil f \rceil_{x^\circ}$ the corresponding linearisation error. After each iteration of the algorithm, the guessed trajectories are updated with the solution of the MPC optimisation and a new pass is initiated by linearising the dynamics around the new estimate. 

\subsubsection{Parameterisation of the uncertainty sets} 
\begin{figure}
     \centering
     \begin{subfigure}[b]{0.45\textwidth}
         \centering
         \includegraphics[width=\textwidth, trim={5cm 5cm 4.5cm 6cm}, clip]{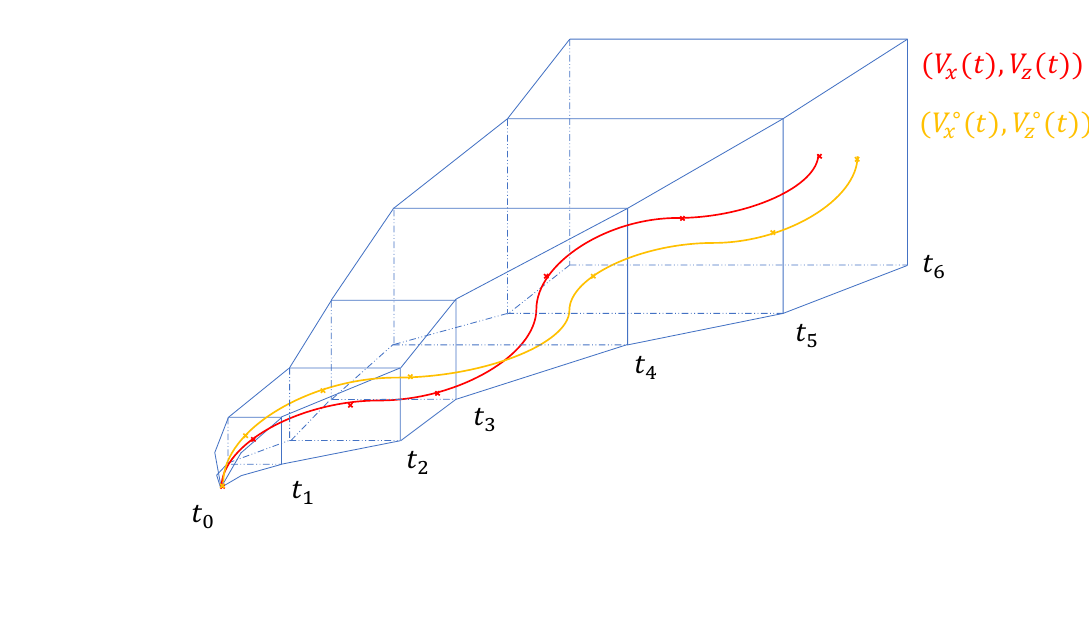}
         \caption{Tube}
         \label{fig:tube1}
     \end{subfigure}
     \hfill
     \begin{subfigure}[b]{0.45\textwidth}
         \centering
         \includegraphics[width=\textwidth, trim={5cm 5cm 4.5cm 6cm}, clip]{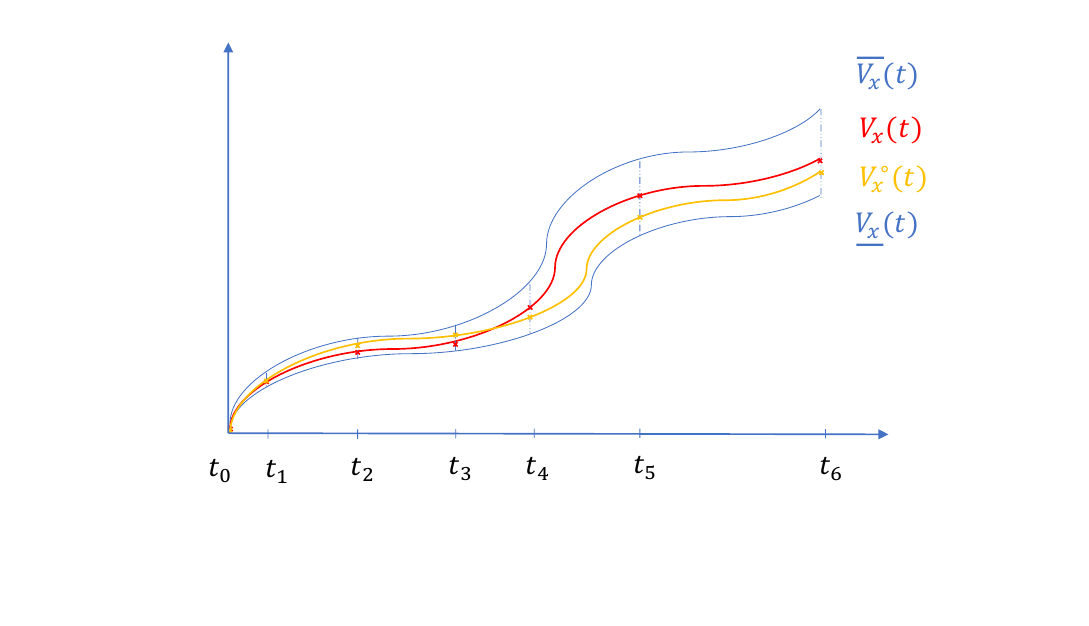}
         \caption{Tube for $V_x$ only}
         \label{fig:tube3}
     \end{subfigure}
    \hfill
     \begin{subfigure}[b]{0.45\textwidth}
         \centering
         \includegraphics[width=\textwidth, trim={7cm 7cm 7cm 6.5cm}, clip]{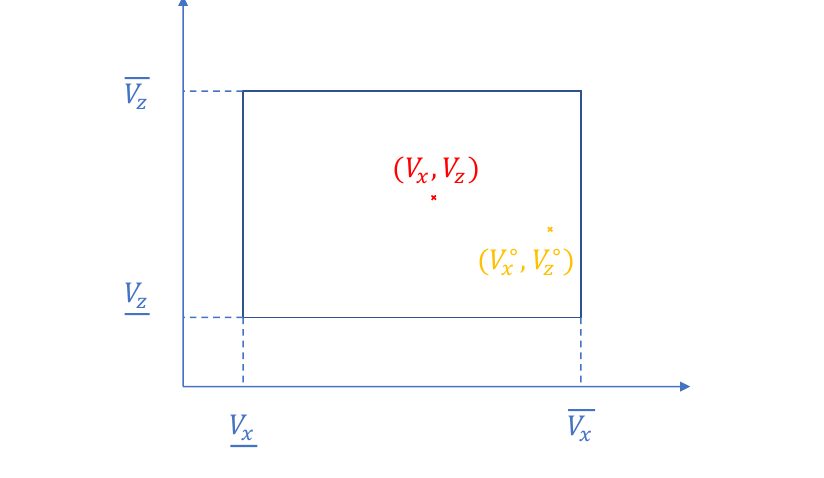}
         \caption{Tube cross section}
         \label{fig:tube2}
     \end{subfigure}
        \caption{Schematic visualisation of the tube. The uncertain state trajectory (red) lies within a tube (blue) centred around the nominal trajectory (orange), sampled at various time steps (\ref{fig:tube1}). Also shown is the tube evolution for state $V_x$ only (\ref{fig:tube3}) and a snapshot of the tube cross section at an arbitrary time (\ref{fig:tube2}).}
        \label{fig:tube}
\end{figure}

We assume that the uncertain state trajectories $V_x, V_z$ lie within
``tubes'' whose cross-sections are parameterised by means of elementwise bounds $V_x \in [\underline{V}_x, \overline{V}_x], V_z \in [\underline{V}_z, \overline{V}_z]$, which are optimisation variables, see Figure \ref{fig:tube}. 

\subsubsection{DC properties of dynamics} By convexity of $g_1, g_2, h_1, h_2$, the associated linearisation errors are necessarily convex and take their maximum on the boundary of the set over which the functions are constrained. Moreover, by definition, their minimum on this set is zero (Jacobian linearisation). It follows that the bounds on the states dynamics satisfy the following convex inequalities
\begin{gather}
\label{eq:tube1}
m\dot{\overline{V_x}} \geq \max_{\substack{V_x \in \{\underline{V}_x, \overline{V}_x\}, V_z \in \{\underline{V}_z, \overline{V}_z\}}} \bigl\{ g_1(V_x, V_z, \mu, \tau) - \lfloor h_1 \rfloor_{(V_x^\circ, V_z^\circ, \mu^\circ, \tau^\circ)} (V_x, V_z, \mu, \tau )+ \overline{W}_x \bigr\}, \\
\label{eq:tube2}
m\dot{\overline{V_z}} \geq \max_{\substack{V_x \in \{\underline{V}_x, \overline{V}_x\}, V_z \in \{\underline{V}_z, \overline{V}_z\}}} \bigl\{ g_2(V_x, V_z, \mu, \tau) - \lfloor h_2 \rfloor_{(V_x^\circ, V_z^\circ, \mu^\circ, \tau^\circ)} (V_x, V_z, \mu, \tau )+ \overline{W}_z \bigr\},  \\
m\dot{\underline{V_x}} \leq \min_{\substack{V_x \in \{\underline{V}_x, \overline{V}_x\}, V_z \in \{\underline{V}_z, \overline{V}_z\}}} \bigl\{ -h_1(V_x, V_z, \mu, \tau) + \lfloor g_1 \rfloor_{(V_x^\circ, V_z^\circ, \mu^\circ, \tau^\circ)} (V_x, V_z, \mu, \tau )+ \underline{W}_x \bigr\},\\
%\label{eq:tube3}
m\dot{\underline{V_z}} \leq \min_{\substack{V_x \in \{\underline{V}_x, \overline{V}_x\}, V_z \in \{\underline{V}_z, \overline{V}_z\}}} \bigl\{ -h_2(V_x, V_z, \mu, \tau) + \lfloor g_2 \rfloor_{(V_x^\circ, V_z^\circ, \mu^\circ, \tau^\circ)} (V_x, V_z, \mu, \tau )+ \underline{W}_z \bigr\},
\label{eq:tube4}
\end{gather}
Conditions \eqref{eq:tube1}-\eqref{eq:tube4} must be satisfied by the tube bounding uncertain model trajectories.
%define the tube for the nonlinear part of the dynamics. 
These constraints involve only minimisations of linear functions and maximisations of convex functions. 
Therefore they reduce to a finite number of constraints involving the tube vertices (i.e the variables
%The functions to optimise no longer need to be evaluated on continuous intervals but on their vertices 
$\{\underline{V}_x, \overline{V}_x\}, \{\underline{V}_z, \overline{V}_z\}$). Thus each of the constraints \eqref{eq:tube1}-\eqref{eq:tube4}
%maximisation and minimisation operation 
reduces to $2^2 = 4$ convex inequalities.\\

\subsection{Discrete time DC-TMPC}
\label{sub:discrete}

In order to obtain a finite dimensional robust MPC optimisation, we discretise the problem with a fixed sampling interval $\delta$ and evaluate all variables over a finite horizon $N$. The notation $\{x_0, x_1, \dots,  x_{N-1}\}$ is used for the sequence of current and future values of a variable $x$ predicted at the $n$-th discrete-time step, so that $x_k$ denotes the predicted value of $x((n + k)\delta)$.

The MPC optimisation at the $n$-th discrete-time step is initialised with a feasible predicted trajectory $(V_{x}^\circ, V_{z, k}^\circ, \mu_k^\circ,  \tau_k^\circ)$ and the following optimisation problem $\mathcal{P}$ (obtained by discretising and gathering equations \eqref{eq:second_order}-\eqref{eq:obj}, \eqref{eq:tube1}-\eqref{eq:tube4}) is solved sequentially 
%
%\begin{equation}
\begin{alignat}{3}
& \min_{\substack{\overline{V}_{x},\,\underline{V}_{x}, \overline{V}_{z},\,\underline{V}_{z},\\ \tau,\, \mu}} & &  \max_{\substack{V_x \in \{\underline{V}_{x, k}, \overline{V}_{x, k}\}, \\ V_z \in \{\underline{V}_{z, k}, \overline{V}_{z, k}\}}} \Biggl\{\sum_{k=0}^{N-1} \quad \Biggl\| \Biggl[ \begin{matrix}{V}_{x} - V_{x, k}^r \\ {V}_{z} - V_{z, k}^r\end{matrix}  \Biggr] \Biggr\|_{Q_x}^2 + \Biggl\| \Biggl[ \begin{matrix}\mu_{k} + K_{i_{w, k}} (V_{x} - V_{x, k}^\circ) + K_{i_{w, k}}' (V_{z} - V_{z, k}^\circ) - i_{w, k}^r \\ \tau_k + K_{T_k} ({V}_{x} - V_{x, k}^\circ) + K_{T_k}' ({V}_{z} - V_{z, k}^\circ) - T^r_k\end{matrix}  \Biggr] \Biggr\|_{Q_u}^2 \Biggr\}  \nonumber\\ & & & + \max_{\substack{V_x \in \{\underline{V}_{x, N}, \overline{V}_{x, N}\},\\ V_z \in \{\underline{V}_{z, N}, \overline{V}_{z, N}\}}} \Biggl\| \Biggl[ \begin{matrix} {V}_{x} - V_{x, N}^r \\ {V}_{z} - V_{z, N}^r \end{matrix}  \Biggr] \Biggr\|_{\hat{Q}}^2 ,  \nonumber\\ & \text{s.t.},  & &  \forall k \in \{0, \ldots, N-1\}, \nonumber\\
& & & m{\overline{V}_{x, k+1}} \geq \max_{\substack{V_x \in \{\underline{V}_{x, k}, \overline{V}_{x, k}\},\\ V_z \in \{\underline{V}_{z, k}, \overline{V}_{z, k}\}}} \bigl\{ m V_{x} + \delta g_1(V_x, V_z, \mu_k, \tau_k) - \delta\lfloor h_1 \rfloor_{(V_{x, k}^\circ, V_{z, k}^\circ, \mu_k^\circ, \tau_k^\circ)} (V_x, V_z, \mu_k, \tau_k)+ \delta \overline{W}_{x, k} \bigr\},
\nonumber\\
& & & m{\overline{V}_{z, k+1}} \geq \max_{\substack{V_x \in \{\underline{V}_{x, k}, \overline{V}_{x, k}\},\\ V_z \in \{\underline{V}_{z, k}, \overline{V}_{z, k}\}}} \bigl\{ m V_{z} + \delta g_2(V_x, V_z, \mu_k, \tau_k) - \delta\lfloor h_2 \rfloor_{(V_{x, k}^\circ, V_{z, k}^\circ,  \mu_k^\circ, \tau_k^\circ)} (V_x, V_z,  \mu_k, \tau_k)+ \delta \overline{W}_{z, k} \bigr\},
\nonumber\\
& & & m{\underline{V}_{x, k+1}} \leq \min_{\substack{V_x \in \{\underline{V}_{x, k}, \overline{V}_{x, k}\},\\ V_z \in \{\underline{V}_{z, k}, \overline{V}_{z, k}\}}} \bigl\{ m V_{x} - \delta h_1(V_x, V_z,  \mu_k, \tau_k) + \delta\lfloor g_1 \rfloor_{(V_{x, k}^\circ, V_{z, k}^\circ,  \mu_k^\circ, \tau_k^\circ)} (V_x, V_z,  \mu_k, \tau_k)+ \delta \underline{W}_{x, k} \bigr\},
\nonumber\\
& & & m{\underline{V}_{z, k+1}} \leq \min_{\substack{V_x \in \{\underline{V}_{x, k}, \overline{V}_{x, k}\},\\ V_z \in \{\underline{V}_{z, k}, \overline{V}_{z, k}\}}} \bigl\{ m V_{z} - \delta h_2(V_x, V_z,  \mu_k, \tau_k) + \delta\lfloor g_2 \rfloor_{(V_{x, k}^\circ, V_{z, k}^\circ,  \mu_k^\circ, \tau_k^\circ)} (V_x, V_z,  \mu_k, \tau_k)+ \delta \underline{W}_{z, k} \bigr\},
\nonumber\\
& & &  \underline{V}_{x, 0} = \overline{V}_{x, 0} = V_x(n\delta), \, \underline{V}_{z, 0} = \overline{V}_{z, 0} = V_z(n\delta),
\nonumber \\
& & &  \underline{i}_w \leq \min_{\substack{V_x \in \{\underline{V}_{x, k}, \overline{V}_{x, k}\},\\ V_z \in \{\underline{V}_{z, k}, \overline{V}_{z, k}\}}} \bigl\{\mu_{k} + K_{i_{w, k}} (V_{x} - V_{x, k}^\circ) + K_{i_{w, k}}' (V_{z} - V_{z, k}^\circ) \bigr\} 
\nonumber \\
& & & \overline{i}_w \geq \max_{\substack{V_x \in \{\underline{V}_{x, k}, \overline{V}_{x, k}\},\\ V_z \in \{\underline{V}_{z, k}, \overline{V}_{z, k}\}}} \bigl\{\mu_{k} + K_{i_{w, k}} (V_{x} - V_{x, k}^\circ) + K_{i_{w, k}}' (V_{z} - V_{z, k}^\circ) \bigr\}
\nonumber\\
& & & \underline{V}_x \leq \underline{V}_{x, k} , \, \overline{V}_{x, k} \leq \overline{V}_x, \,  \underline{V}_z \leq \underline{V}_{z, k} , \,  \overline{V}_{z, k} \leq \overline{V}_z,
\nonumber\\
& & &  \underline{a} \leq \frac{ \underline{V}_{x, k+1} - \overline{V}_{x, k}}{\delta} , \, \underline{a} \leq \frac{\underline{V}_{z, k+1} - \overline{V}_{z, k}}{\delta} , \,    \frac{ \overline{V}_{x, k+1} - \underline{V}_{x, k}}{\delta} \leq \overline{a} , \, \frac{\overline{V}_{z, k+1} - \underline{V}_{z, k}}{\delta} \leq \overline{a},
\nonumber\\
& & & 0 \leq \min_{\substack{V_x \in \{\underline{V}_{x, k}, \overline{V}_{x, k}\},\\ V_z \in \{\underline{V}_{z, k}, \overline{V}_{z, k}\}}} \bigl\{ \tau_k + K_{T_k} ({V}_{x} - V_{x, k}^\circ) + K_{T_k}' ({V}_{z} - V_{z, k}^\circ) \bigr\},  
\nonumber\\ 
& & & \overline{T}\geq \max_{\substack{V_x \in \{\underline{V}_{x, k}, \overline{V}_{x, k}\},\\ V_z \in \{\underline{V}_{z, k}, \overline{V}_{z, k}\}}} \bigl\{ \tau_k + K_{T_k} ({V}_{x} - V_{x, k}^\circ) + K_{T_k}' ({V}_{z} - V_{z, k}^\circ) \bigr\} , 
\nonumber\\
& & & \underline{M} \leq \min_{\substack{V_x \in \{\underline{V}_{x, k}, \overline{V}_{x, k}\},\\ V_z \in \{\underline{V}_{z, k}, \overline{V}_{z, k}\}}} \bigl\{ J_w \Delta^2 \{\mu_{k} + K_{i_{w, k}} (V_{x, k} - V_{x, k}^\circ) + K_{i_{w, k}}' (V_{z, k} - V_{z, k}^\circ)\} \bigr\},
\nonumber\\
& & & \overline{M} \geq \max_{\substack{V_x \in \{\underline{V}_{x, k}, \overline{V}_{x, k}\},\\ V_z \in \{\underline{V}_{z, k}, \overline{V}_{z, k}\}}} \bigl\{ J_w \Delta^2 \{\mu_{k} + K_{i_{w, k}} (V_{x, k} - V_{x, k}^\circ) + K_{i_{w, k}}' (V_{z, k} - V_{z, k}^\circ)\} \bigr\},
\nonumber\\
& & & \hat{\gamma} \geq \max_{\substack{V_x \in \{\underline{V}_{x, N}, \overline{V}_{x, N}\},\\ V_z \in \{\underline{V}_{z, N}, \overline{V}_{z, N}\}}} \Biggl\| \Biggl[ \begin{matrix} {V}_{x} - V_{x, N}^r \\ {V}_{z} - V_{z, N}^r \end{matrix} \Biggr] \Biggr\|  _{\hat{Q}}^2 
\end{alignat}
%\end{equation}
where $V_x(0), V_z(0)$ at time step $n = 0$ are given in Table \ref{tab:BC} depending on the transition scenario considered. We defined the second order forward finite difference operator as $\Delta^2 f_k = \frac{f_{k+2} - 2 f_{k+1} + f_k}{\delta^2} $. Note that the possible vertices for the tube are given by
\[ \mathcal{V} = \left\{ \begin{bmatrix} \underline{V}_{x, k}\\ \underline{V}_{z, k} \end{bmatrix}, \begin{bmatrix} \overline{V}_{x, k}\\ \underline{V}_{z, k}  \end{bmatrix}, \begin{bmatrix} \underline{V}_{x, k}\\ \overline{V}_{z, k} \end{bmatrix}, \begin{bmatrix} \overline{V}_{x, k}\\ \overline{V}_{z, k}  \end{bmatrix} \right\}, 
\] 
which allows us to express each maximisation / minimisation above as a set of 4 inequalities at most. Moreover, since the feedback gains and the terminal penalty matrix are known \textit{a priori}, this number can be further reduced to 2 for all inequalities but the first four.  

Once problem $\mathcal{P}$ is solved, the guessed trajectories are updated with the solution as follows \cite{DC-TMPC} 
\begin{gather}
\label{eq:update1}
{V}_{x, 0} \leftarrow V_x(n\delta) , \quad  {V}_{z, 0} \leftarrow V_z(n\delta) , \\
i_{w, k}^\circ \leftarrow  \mu_k +  K_{i_{w, k}} (V_{x, k} - V_{x, k}^\circ)  +  K'_{i_{w, k}} ({V}_{z, k} - V_{z, k}^\circ),\\ 
T_k^\circ \leftarrow \tau_k + K_{T_k} (V_{x, k} - V_{x, k}^\circ) + K_{T_k}' ({V}_{z, k} - V_{z, k}^\circ), \\
V_{x, k+1} \leftarrow V_{x, k} + \delta f_1 (V_{x, k}, V_{z, k}, i_{w, k}^\circ,  T_k^\circ)/m, \\
V_{z, k+1} \leftarrow V_{z, k} + \delta f_2 (V_{x, k}, V_{z, k}, i_{w, k}^\circ,  T_k^\circ)/m, \\
V_{x, k}^\circ \leftarrow V_{x, k+1} , \quad V_{z, k}^\circ \leftarrow V_{z, k+1}, \\
\mu_{k}^\circ \leftarrow \mu_{k} , \quad \tau_{k}^\circ \leftarrow \tau_{k},
\label{eq:update7}
\end{gather}
for $k = 0, \dots N-1$ and the process of solving $\mathcal{P}$ and updating the trajectories with \eqref{eq:update1}-\eqref{eq:update7} is repeated until $|| [\tau \quad \mu]^\top || < \epsilon$. The control law at time $n$ is then implemented by taking the first element of the control sequence
\[
i_w(n\delta) = i_{w, 0}^\circ, \quad T(n\delta) = T_0^\circ.
\]
At time $n+1$, we set $V_{x, 0} = V_{x}((n+1)\delta)$, $V_{z, 0} = V_{z}((n+1)\delta)$, and update, $\forall k = 0, \dots, N-2$ \cite{DC-TMPC}
\begin{gather}
i_{w, k}^\circ \leftarrow i_{w, k+1}^\circ, \, T_{k}^\circ \leftarrow T_{k+1}^\circ, \, \mu_{k}^\circ \leftarrow \mu_{k+1}^\circ, \, \tau_{k}^\circ \leftarrow \tau_{k+1}^\circ,\\
V_{x, k+1}^\circ \leftarrow V_{x, k}^\circ + \delta (f_1 (V_{x, k}^\circ, V_{z, k}^\circ, i_{w, k}^\circ,  T_k^\circ) + W_x)/m,\\
V_{z, k+1}^\circ \leftarrow V_{z, k}^\circ + \delta (f_2 (V_{x, k}^\circ, V_{z, k}^\circ, i_{w, k}^\circ,  T_k^\circ) + W_z)/m,
\end{gather}
and finally, as per the dual mode MPC paradigm \cite{DC-TMPC}
\begin{gather}
i_{w, N-1}^\circ \leftarrow  \hat{K}_{i_w} (V_{x, N-1}^\circ - V_{x, N-1}^r) + V_{x, N-1}^r +  \hat{K}_{i_w}' (V_{z, N-1}^\circ - V_{z, N-1}^r) + V_{z, N-1}^r, \\
T_{N-1}^\circ \leftarrow  \hat{K}_{T} (V_{x, N-1}^\circ - V_{x, N-1}^r) + V_{x, N-1}^r +  \hat{K}_{T}' (V_{z, N-1}^\circ - V_{z, N-1}^r) + V_{z, N-1}^r,\\
V_{x, N}^\circ \leftarrow V_{x, N-1}^\circ + \delta (f_1 (V_{x, N-1}^\circ, V_{z, N-1}^\circ, i_{w, N-1}^\circ,  T_{N-1}^\circ) + W_x)/m, \\
V_{z, N}^\circ \leftarrow V_{z, N-1}^\circ + \delta (f_2 (V_{x, N-1}^\circ, V_{z, N-1}^\circ, i_{w, N-1}^\circ,  T_{N-1}^\circ) + W_z)/m,
\end{gather}
where the terminal gains $\hat{K}_{i_w}, \hat{K}_{i_w}', \hat{K}_{T}, \hat{K}_{T}'$ can be computed following the Appendix in \cite{DC-TMPC}.

\section{Results}
\label{sec:results}
We consider a case study based on
%inspired from 
the transition of the Airbus A$^3$ Vahana (i) from powered to wing-borne flight (forward transition) and (ii) from wing-borne to powered flight (backward transition). In what follows, unless otherwise stated, simulations are conducted in the absence of wind. Parameters and transition boundary conditions are reported in Table \ref{tab:param} and \ref{tab:BC}. The terminal times for the forward and backward transitions are respectively set to $t_f=25 s$ and $t_f = 17 s$  and the time step is $\delta \approx 0.22 s$ in both cases, resulting in respectively $N=110$ and $N=75$ discretisation points. Optimisation problem $\mathcal{P}$ is solved using CVXPY \cite{cvxpy} with solver MOSEK \cite{mosek}.

\begin{table}[ht]
\centering
\begin{tabular}{llll}
\hline
\textbf{Parameter} & \textbf{Symbol} & \textbf{Value} & \textbf{Units} \\ \hline
Mass     &    $m$    &   $752.2$    &    \si{kg}   \\ \hline
Gravity acceleration     &    $g$    &   $9.81$    &    \si{m.s^{-2}}   \\ \hline
Wing area &   $S$     &    $8.93$   &    \si{m^2}   \\ \hline
Disk area &   $A$     &    $2.83$   &    \si{m^2} \\ \hline
Wing inertia &   $J_w$     &    $1100$   &    \si{kg.m^2}   \\ \hline
%Induced speed coefficient &   $k_w$     &    $2$   &    \si{-} \\ \hline
Density of air &   $\rho$     &    $1.225$   &    \si{kg.m^{-3}} \\ \hline
Maximum thrust &   $\overline{T}$     &    $8855$   &    \si{N} \\ \hline
%Angle of attack range &   $\left[\underline{\alpha}, \overline{\alpha}\right]$     &    $\left[-10, 60\right]$   &    \si{deg}   \\ \hline
%Flight path angle range &   $\left[\underline{\gamma}, \overline{\gamma}\right]$     &    $\left[-90, 90\right]$   &    \si{deg}   \\ \hline
Tiltwing angle range &   $\left[\underline{i}_w, \overline{i}_w\right]$     &    $\left[-10, 100\right]$   &    \si{deg}   \\ \hline
Acceleration range &   $\left[\underline{a}, \overline{a}\right]$     &    $\left[-0.3g, 0.3g\right]$   &    \si{m . s^{-2}}   \\ \hline
Forward velocity range &   $\left[\underline{V}_x, \overline{V}_x\right]$     &    $\left[0, 60\right]$   &    \si{m/s}   \\ \hline
Vertical velocity range &   $\left[\underline{V}_z, \overline{V}_z\right]$     &    $\left[-30, 30\right]$   &    \si{m/s}   \\ \hline
Torque range &   $\left[\underline{M}, \overline{M}\right]$     &    $\left[-50; 50\right]$   &    \si{N.m}   \\ \hline
%Velocity boundary conditions &   $\left\{V_0; V_f\right\}$     &    $\left\{1, 40\right\}$   &    \si{m/s}   \\ \hline
Number of propellers &   $n$     &    $4$   &    \si{-} \\ \hline
Time step &   $\delta$     &    $0.22$   &    \si{s} \\ \hline 
Degree of polynomial $f$ &   $2d$     &    $2$   &    \si{-} \\ \hline 
\end{tabular}
\vspace{1mm}\caption{Model parameters derived from A$^3$ Vahana}
\label{tab:param}
\vspace{-3mm}
\end{table}
\begin{table}[ht]
\centering
\begin{tabular}{llll}
\hline
\textbf{Parameter} & \textbf{Symbol} & \textbf{Value} & \textbf{Units} \\ \hline
\multicolumn{4}{c}{{\cellcolor[rgb]{0.753,0.753,0.753}}\textbf{Forward transition}}    \\
Forward velocity &   $\left\{V_{x, 0}; V_{x, f}\right\}$  &    $\left\{0; 40\right\}$ &    \si{m/s}   \\ \hline
Vertical velocity  &  $\left\{V_{z, 0}; V_{z, f}\right\}$ &    $\left\{0; 0\right\}$   &    \si{m/s}   \\ \hline
%Tiltwing angle  &   $i_0$     &    $75$   &    \si{deg}   \\ \hline
\multicolumn{4}{c}{{\cellcolor[rgb]{0.753,0.753,0.753}}\textbf{Backward transition}}    \\
Forward velocity &   $\left\{V_{x, 0}; V_{x, f}\right\}$  &    $\left\{40; 0\right\}$ &    \si{m/s}   \\ \hline
Vertical velocity  &  $\left\{V_{z, 0}; V_{z, f}\right\}$ &    $\left\{0; 0\right\}$   &    \si{m/s}   \\ \hline
%Tiltwing angle  &   $i_0$     &    $0$    &    \si{deg}   \\ \hline
\end{tabular}
\vspace{1mm}\caption{Boundary conditions for transitions}
\label{tab:BC}
\vspace{-3mm}
\end{table}

\subsection{DC decomposition}

The DC decompositions of $f_1$ and $f_2$ are computed according Section \ref{sub:DC}. In each case, the approximation polynomial degree $2d$ is set to $2$ and the nonlinear model $f$ is sampled at $N_s = 1\mathrm{e}{+4}$ evaluation points.   $500$ random test points are then generated to obtain the results presented in Table \ref{tab:DC}. 

\begin{table}[ht]
\centering
\begin{tabular}{llll}
\hline
\textbf{Function} & \textbf{LS mean relative error ($\%$)} & \textbf{Residue of $y^\top(Q - R - P)y =0$} & \textbf{Occurence of non PSD Hessian} \\ \hline
$f_1$ &   $5.8$  &    $5\mathrm{e}{-15}$ &    None   \\ \hline
$f_2$  &  $7.5$ &    $2\mathrm{e}{-12}$   &   None   \\ \hline
\end{tabular}
\vspace{1mm}\caption{DC decomposition and least-squares (LS) fit results for $500$ random test points.}
\label{tab:DC}
\vspace{-3mm}
\end{table}

\begin{figure}[ht]
    \centering
    \includegraphics[width=0.7\textwidth]{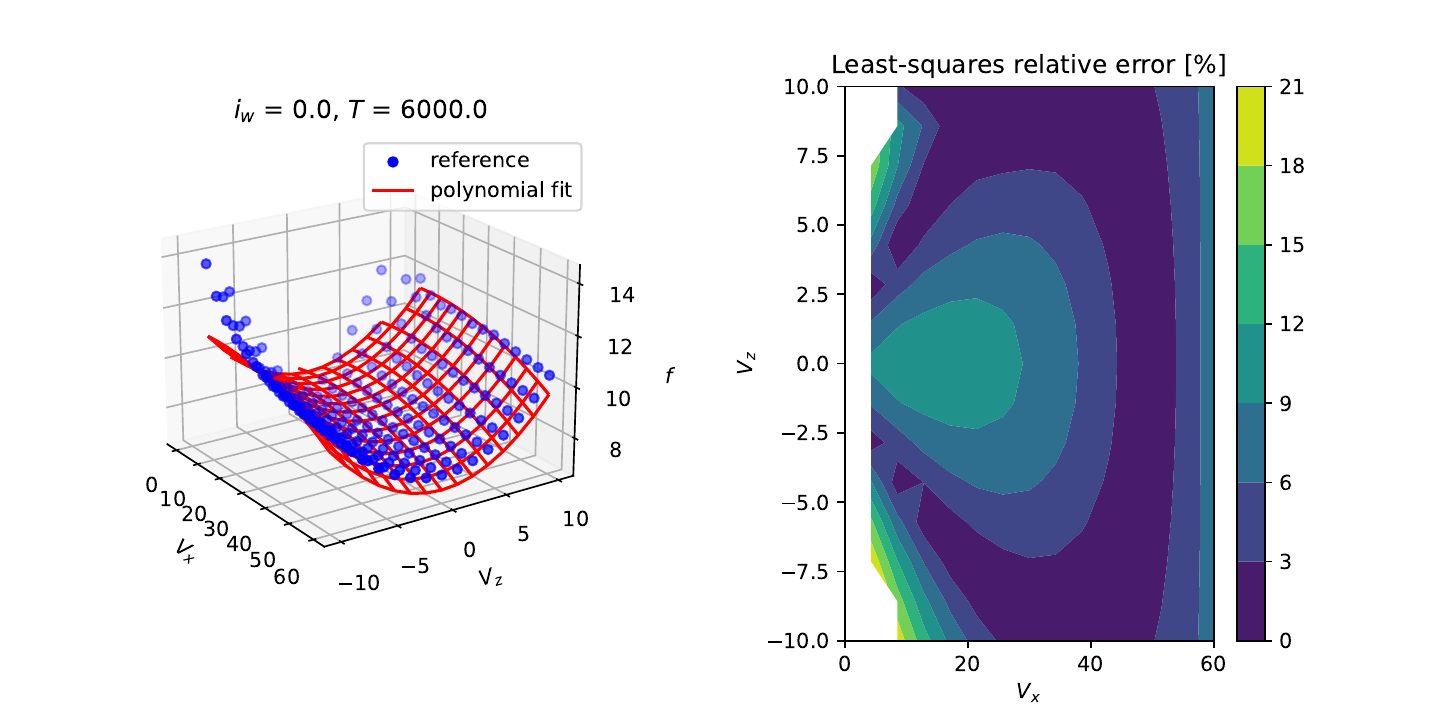} %{<left> <lower> <right> <upper>}
    \caption{Left: least-squares fit of $f_1$ samples (blue dots) by the polynomial model (red curve) given $i_w$ and $T$. Right: contour plot of the percent relative fitting error.}
    \label{fig:fit}
\end{figure}

The least-squares mean relative error measures how well the polynomial model $y^\top P y$ fits the nonlinear model $f$. The obtained errors are acceptable in the present scenario but could be further reduced if increasing the polynomial degree or using a different approximation model (e.g. radial basis functions, neural networks). This would typically come at the cost of increased computation times for the MPC optimisation problem. Figure \ref{fig:fit} illustrates the quality of the fit  for a given tiltwing angle and thrust magnitude (projection is required for visualisation purposes).

The residue of $y^\top(Q - R - P)y =0$ illustrates the accuracy of the DC decomposition of the polynomial approximation, and is excellent in both cases. 

\begin{figure}[ht]
    \centering
    \includegraphics[width=0.7\textwidth]{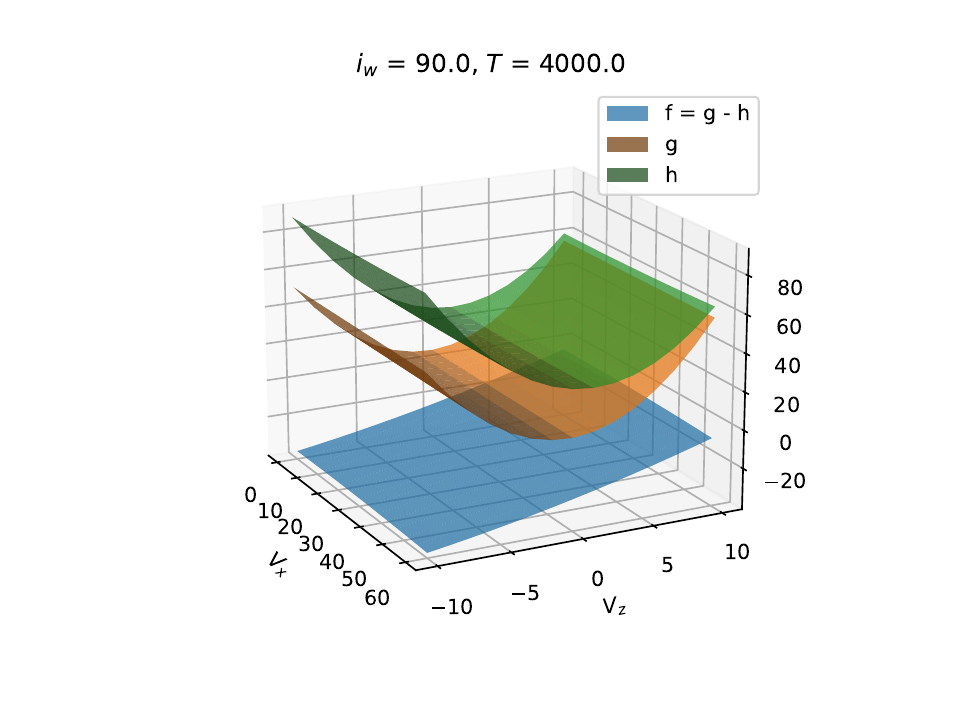} %{<left> <lower> <right> <upper>}
    \caption{DC decomposition $f=g-h$ for given $i_w$ and $T$.}
    \label{fig:DC}
\end{figure}

Finally, we verify that there was no convexity violation by computing the Hessians of the functions at each test point and checking for positive semidefiniteness. A typical DC decomposition is shown in Figure \ref{fig:DC} (with projection).

\subsection{Forward transition}

\begin{figure}[ht]
    \centering
    \includegraphics[width=0.6\textwidth]{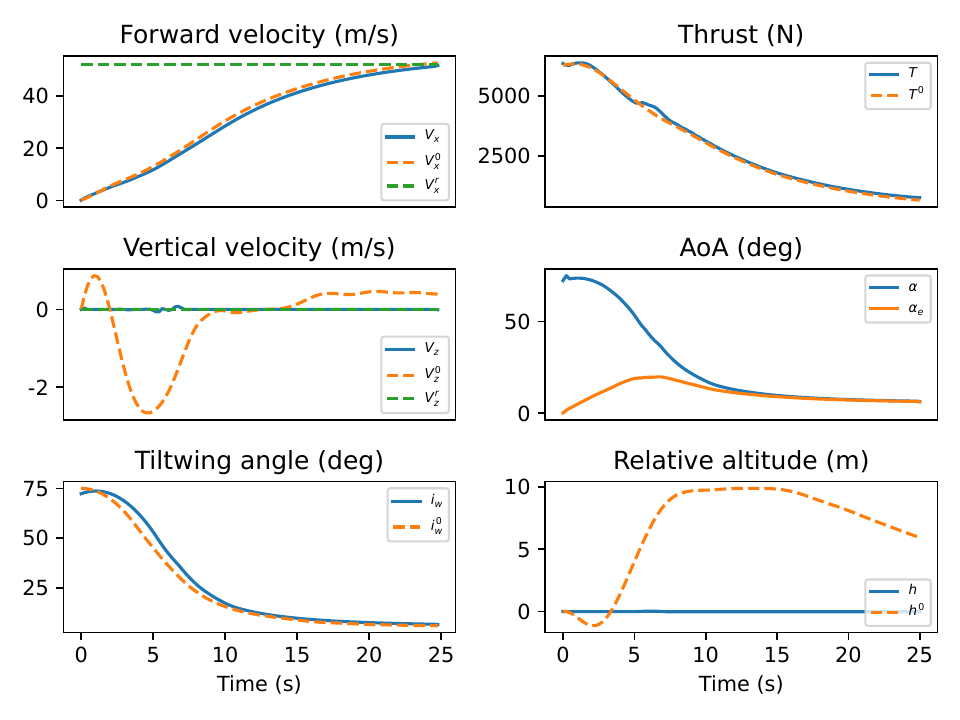} %{<left> <lower> <right> <upper>}
    \caption{Constant altitude forward transition.}
    \label{fig:traj0}
\end{figure}

At first, we set the penalty matrix in the objective to $Q_x = \text{diag}(1, 1\mathrm{e}{+4})$ to achieve a constant altitude forward transition. The results are shown in Figure \ref{fig:traj0}. As the aircraft transitions from powered lift to cruise, the velocity magnitude increases, the thrust and tiltwing angle decrease, illustrating the change in lift generation from propellers to wing. The tiltwing angle drop at the beginning results in an increase in the effective angle of attack. Note that the solution (plain blue) has converged to the desired reference trajectory (dashed green) despite the initial discrepancy with the feasible guess trajectory (dashed orange). 

\begin{figure}[ht]
    \centering
    \includegraphics[width=0.6\textwidth]{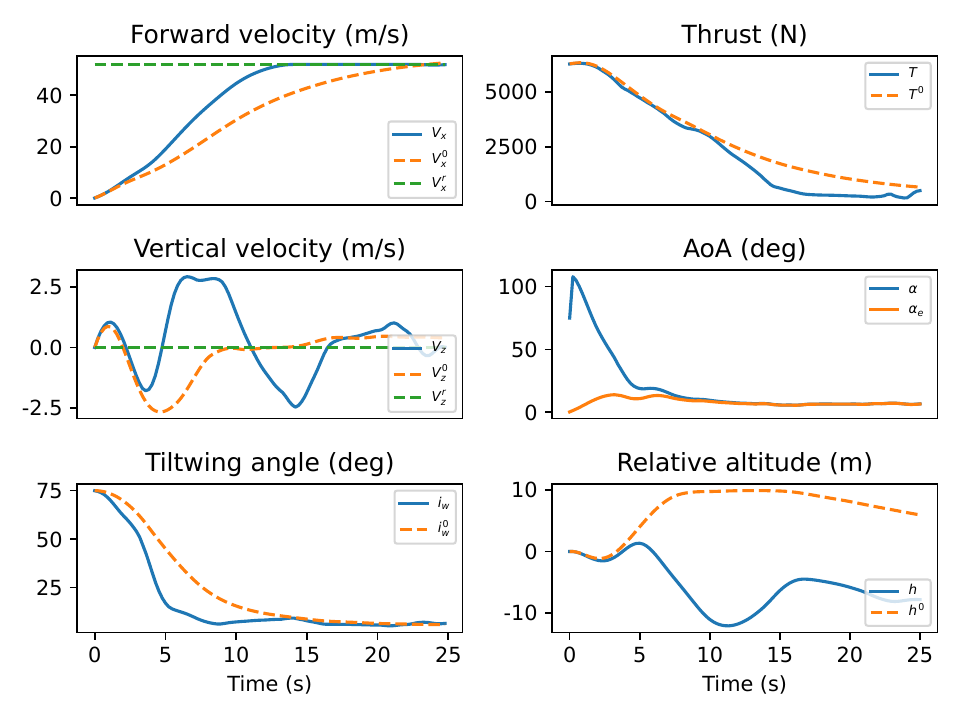} %{<left> <lower> <right> <upper>}
    \caption{A faster forward transition.}
    \label{fig:traj2}
\end{figure}

The objective can be changed to achieve faster transitions. For example, if the penalty matrix is set to $Q_x = \text{diag}(100, 1)$, the obtained results are presented in Figure \ref{fig:traj2}. The reference forward velocity is achieved faster than previously, but this comes at the expense of an altitude drops. A trade-off between both objectives (reaching the desired forward or vertical velocity) can be achieved by varying the penalty matrix. 

\subsection{Backward transition}
\begin{figure}[ht]
    \centering
    \includegraphics[width=0.6\textwidth]{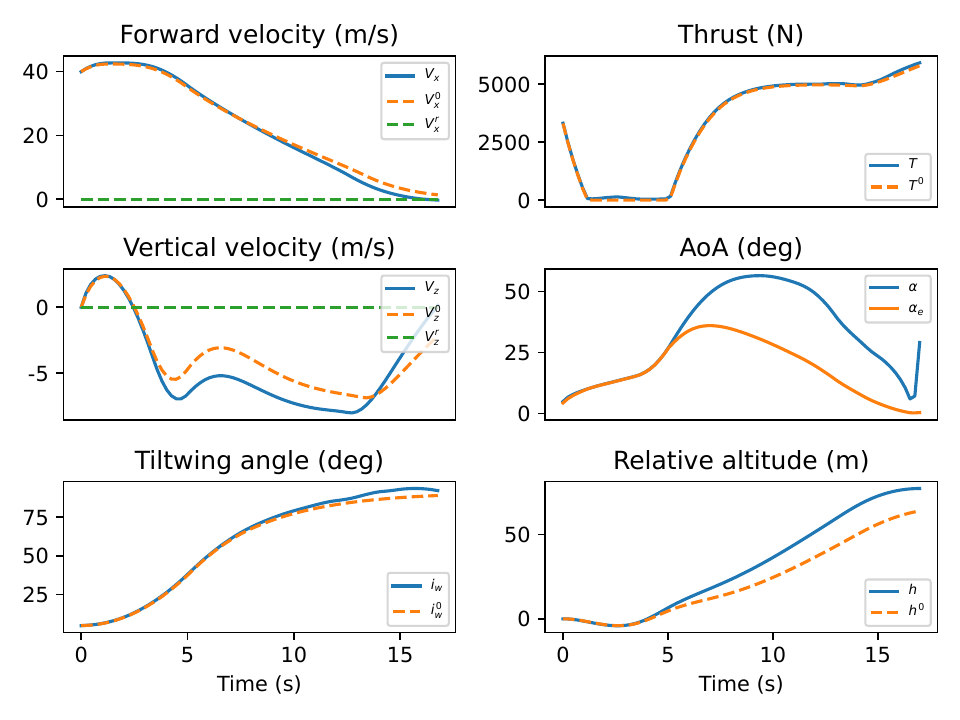} %{<left> <lower> <right> <upper>}
    \caption{Backward transition.}
    \label{fig:traj_back}
\end{figure}

For completeness, we consider the scenario consisting of a backward transition with an increase in altitude, see Figure \ref{fig:traj_back}. This is characterised by a decrease in velocity magnitude and increase in thrust to support the powered flight mode (hover). An increase in altitude of about 75 m is needed for this manoeuvre, and the wing is stalled.

\subsection{Robustness to wind}
\begin{figure}[ht]
    \centering
    \includegraphics[width=0.6\textwidth]{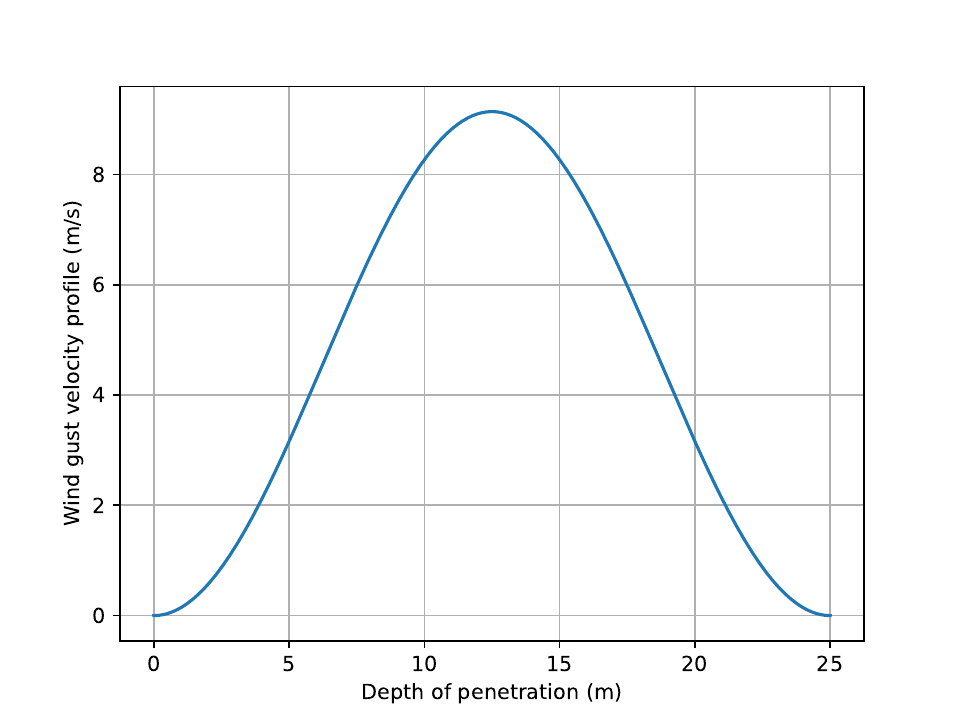} %{<left> <lower> <right> <upper>}
    \caption{Wind gust velocity profile.}
    \label{fig:gust_V}
\end{figure}

To simulate the effect of wind gust on the aircraft, we consider EASA "Means of Compliance with the Special Condition VTOL", §2215 on flight load conditions \cite{EASAvtol} and assume that the aircraft is subject to a discrete wind gust with velocity $U$ following a "one-minus-cosine" law  
\[
U(x_g) =\frac{U_{de}}{2} \left(1 -\cos{\left(\frac{2 \pi x_g}{25 \bar{c}}\right)}\right),
\]
where $0 \leq x_g \leq 25 \bar{c}$ is the distance penetrated into the gust, $\bar{c}$ is the mean geometric chord of the wing, and $U_{de}$ the design gust velocity. The wind gust parameters are reported in Table \ref{tab:gust} and the gust velocity profile with these values is presented in Figure \ref{fig:gust_V}. 

We then consider both crosswind and headwind scenarios for the gust direction. 

\subsubsection{Crosswind}
It is assumed that the wind gust velocity acts normally to the aircraft flight path (velocity vector), i.e. along $\vec{L}$ in Figure \ref{fig:diagram}. This has the effect of modifying the velocity and angle of attack seen by the wing and hence the lift and drag as follows

\begin{equation*}
\label{eq:L1}
L = \tfrac{1}{2}  \lambda \rho S C_L(\alpha_e') V_e'^2 + \tfrac{1}{2} (1-\lambda)  \rho S C_L(\alpha') V'^2, 
\end{equation*}
\begin{equation*}
\label{eq:D1}
D =\tfrac{1}{2} \lambda\rho S C_D(\alpha_e') V_e'^2 + \tfrac{1}{2}  (1-\lambda) \rho S C_D(\alpha') V'^2, 
\end{equation*}

where 
\[
V'^2 = {V^2 + U(x_g)^2}, \quad \alpha' = \alpha + \arctan{\left(\frac{U}{V}\right)},
\]
\[
V_e'^2 = {V_e^2 + U(x_g)^2}, \quad \alpha_e' = \alpha_e + \arctan{\left(\frac{U}{V_e}\right)}. 
\] 

The torque created by the imbalance in lift due to the depth difference along the wing is assumed to be negligible, which justifies our assumption that no wind gust disturbance acts on the tiltwing rotational dynamics in equation \eqref{eq:tiltwing_dyna}.

To evaluate the time varying wind gust bounds $[\underline{W}_i(t), \overline{W}_i(t)]$, $\forall i \in \{x, z\}$ we consider the maximum increment in drag and lift along the guess trajectory as follows
\begin{gather*}
\Delta L_{\text{max}} (t) =  \tfrac{1}{2}  \rho S C_L(\alpha^\circ) U_{de}^2 +  \tfrac{1}{2}  \rho S b_1 \arctan{(U_{de}/V^\circ)} (U_{de}^2 + V^{\circ^2}) \\
\Delta D_{\text{max}} (t) = \tfrac{1}{2}  \rho S C_D(\alpha^\circ) U_{de}^2 +  \tfrac{1}{2}  \rho S ( a_2 (\arctan{(U_{de}/V^\circ)}^2 + 2 \alpha^\circ \arctan{(U_{de}/V^\circ)}) + a_1 \arctan{(U_{de}/V^\circ)} ) (U_{de}^2 + V^{\circ^2}) .
\end{gather*}

\begin{table}[ht]
\centering
\begin{tabular}{llll}
\hline
\textbf{Parameter} & \textbf{Symbol} & \textbf{Value} & \textbf{Units} \\ \hline
Design gust velocity &    $U_{de}$    &   $9.14$    &    \si{m/s}   \\ \hline
Mean geometric chord     &    $\bar{c}$    &   $1$    &    \si{m}   \\ \hline
\end{tabular}
\vspace{1mm}\caption{Wind gust parameters as defined in \cite{EASAvtol}.}
\label{tab:gust}
\vspace{-3mm}
\end{table}

\begin{figure}
     \centering
     \begin{subfigure}[b]{0.33\textwidth}
         \centering
         \includegraphics[width=\textwidth]{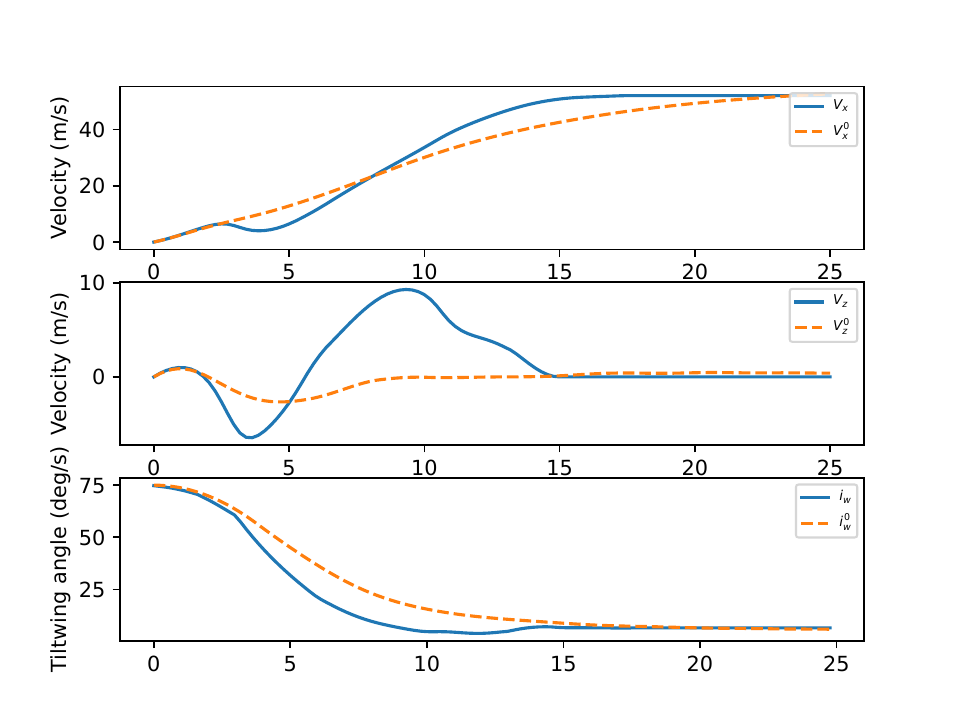}
         \caption{$t=0 s$}
         \label{fig:gust0}
     \end{subfigure}
     \hfill
     \begin{subfigure}[b]{0.33\textwidth}
         \centering
         \includegraphics[width=\textwidth]{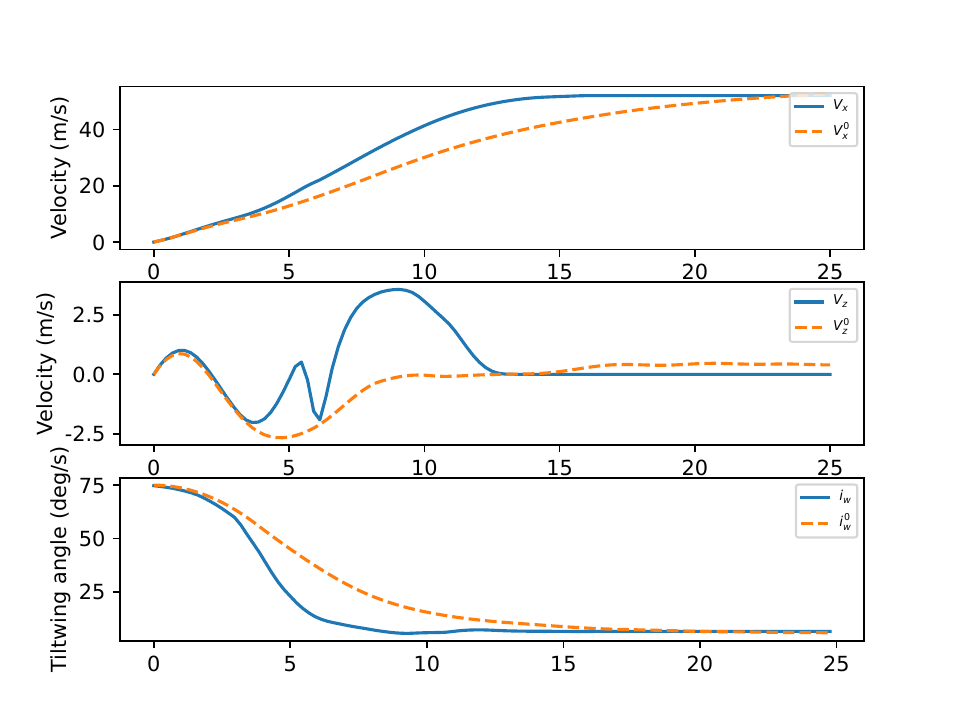}
         \caption{$t=5 s$}
         \label{fig:gust1}
     \end{subfigure}
    \hfill
     \begin{subfigure}[b]{0.33\textwidth}
         \centering
         \includegraphics[width=\textwidth]{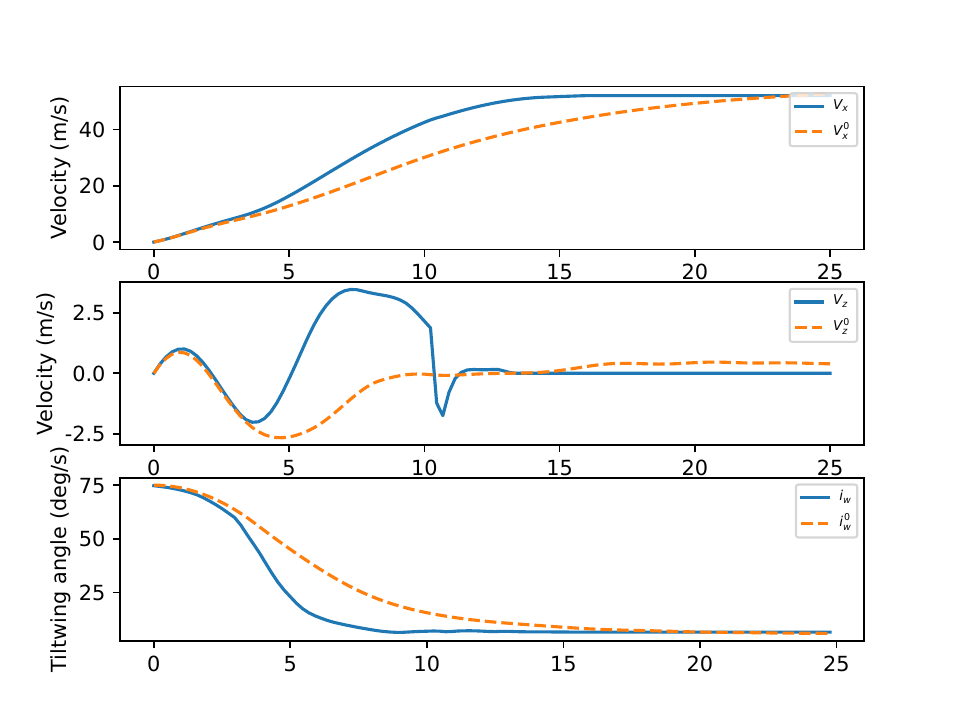}
         \caption{$t=10 s$}
         \label{fig:gust2}
     \end{subfigure}
        \caption{Forward transition with crosswind gust occurring at various times.}
        \label{fig:gust}
\end{figure}

In order to evaluate the effect of wind gusts on the aircraft, we conduct multiple simulations by varying the instant at which the aircraft encounters a wind gust during the forward and backward transitions and we observe the subsequent deviations from the reference: 

\begin{itemize}
    \item \textbf{Forward transition with crosswind.} The results are illustrated in Figure \ref{fig:gust}. For the wind gusts occuring at times $t=5$ $\si{s}$ and $t=10$ $\si{s}$, the deviations observed are reasonable with vertical velocity not exceeding $3$ $\si{m/s}$ in magnitude. The deviation is more important when the disturbance occurs at $t=0$ $\si{s}$ since the vehicle is in hover mode, but the system eventually recovers and stabilises. 
    \item \textbf{Backward transition with crosswind.} The backward transition could not be achieved with crosswind gusts of $9.14 m/s$, so the wind speed was reduced to $5.14 m/s$ to obtain the results in Figure \ref{fig:head}. In all cases, the forward and vertical velocities are slightly perturbed and the system eventually recovers. 
\end{itemize}

\begin{figure}
     \centering
     \begin{subfigure}[b]{0.33\textwidth}
         \centering
         \includegraphics[width=\textwidth]{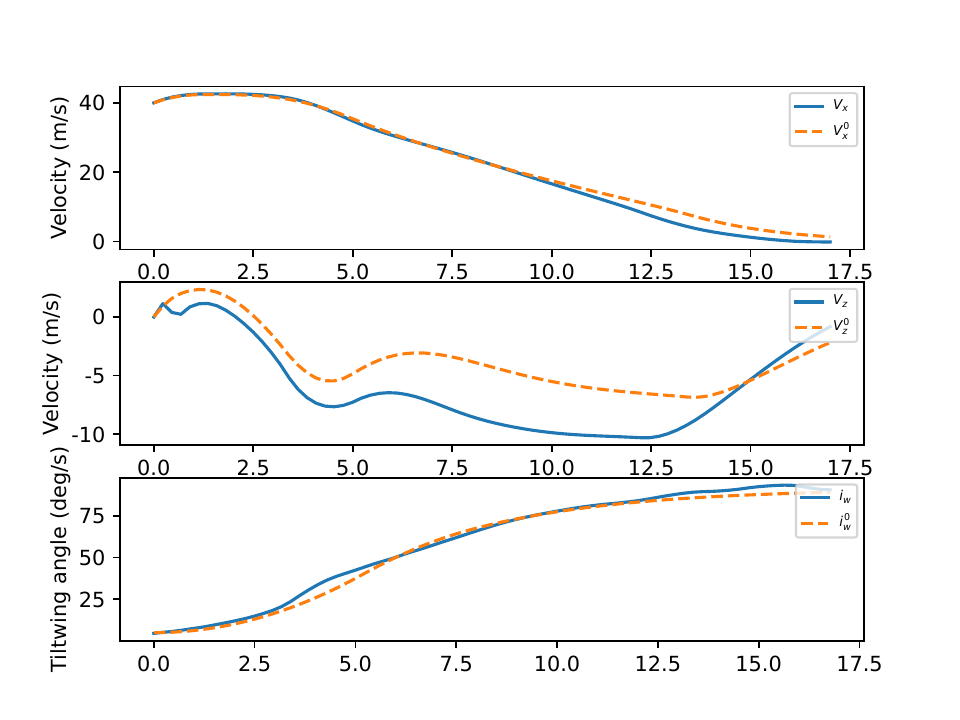}
         \caption{$t=0 s$}
         \label{fig:back0}
     \end{subfigure}
     \hfill
     \begin{subfigure}[b]{0.33\textwidth}
         \centering
         \includegraphics[width=\textwidth]{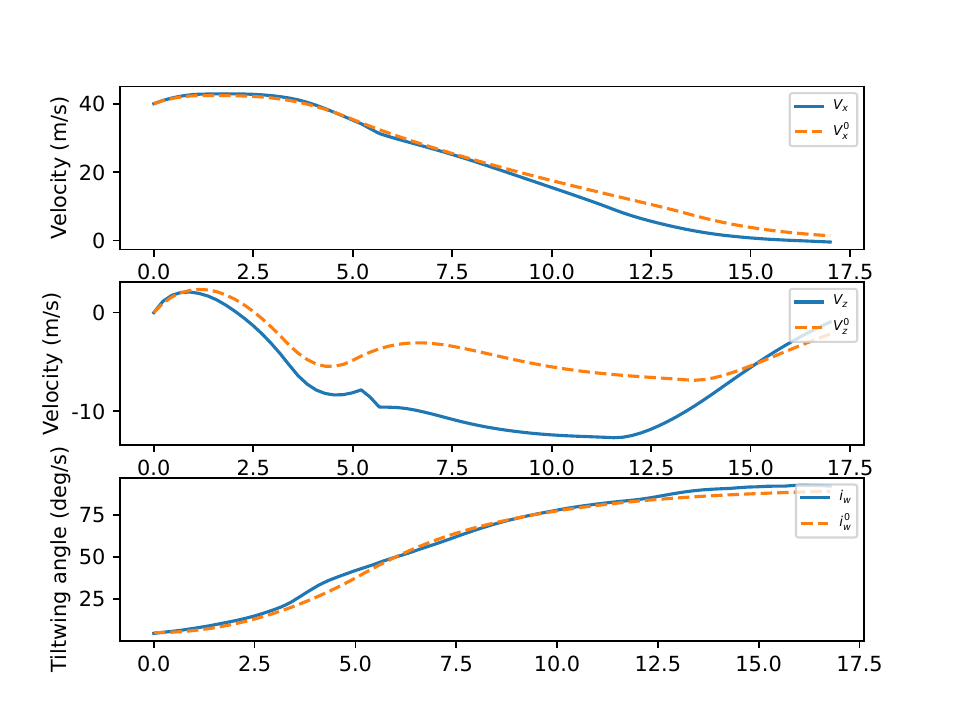}
         \caption{$t=5 s$}
         \label{fig:back1}
     \end{subfigure}
    \hfill
     \begin{subfigure}[b]{0.33\textwidth}
         \centering
         \includegraphics[width=\textwidth]{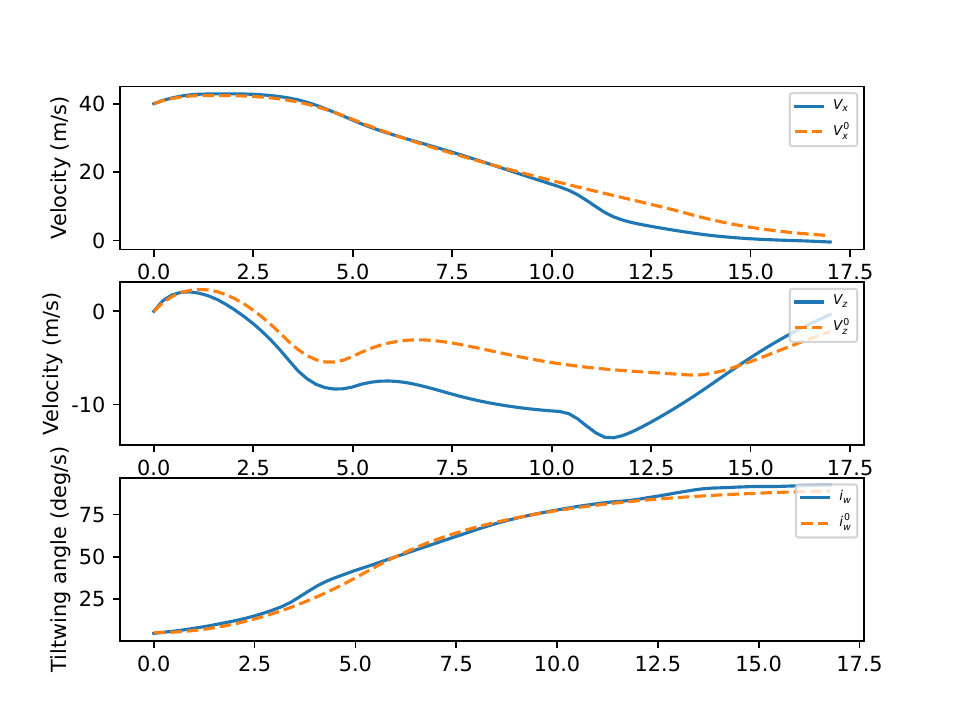}
         \caption{$t=10 s$}
         \label{fig:back2}
     \end{subfigure}
        \caption{Backward transition with crosswind gust occurring at various times.}
        \label{fig:back}
\end{figure}

\subsubsection{Headwind} 
In case of headwind, the wind gust velocity acts anti-parallel to the aircraft velocity vector $\vec{V}$. This modifies the lift and drag as follows (note that this does not affect the angle of attack)

\begin{equation*}
L = \tfrac{1}{2}  \lambda \rho S C_L(\alpha_e) V_e'^2 + \tfrac{1}{2} (1-\lambda)  \rho S C_L(\alpha) V'^2, 
\end{equation*}
\begin{equation*}
D =\tfrac{1}{2} \lambda\rho S C_D(\alpha_e') V_e'^2 + \tfrac{1}{2}  (1-\lambda) \rho S C_D(\alpha') V'^2, 
\end{equation*}

where 
\[
V'^2 = {V^2 + U(x_g)^2},
\]
\[
V_e'^2 = {V_e^2 + U(x_g)^2},
\] 
and we deduce the maximum increment in drag and lift along the guess trajectory as follows
\begin{gather*}
\Delta L_{\text{max}} (t) =  \tfrac{1}{2}  \rho S C_L(\alpha^\circ) U_{de}^2  \\
\Delta D_{\text{max}} (t) = \tfrac{1}{2}  \rho S C_D( \alpha^\circ) U_{de}^2.
\end{gather*}

\begin{figure}
     \centering
     \begin{subfigure}[b]{0.33\textwidth}
         \centering
         \includegraphics[width=\textwidth]{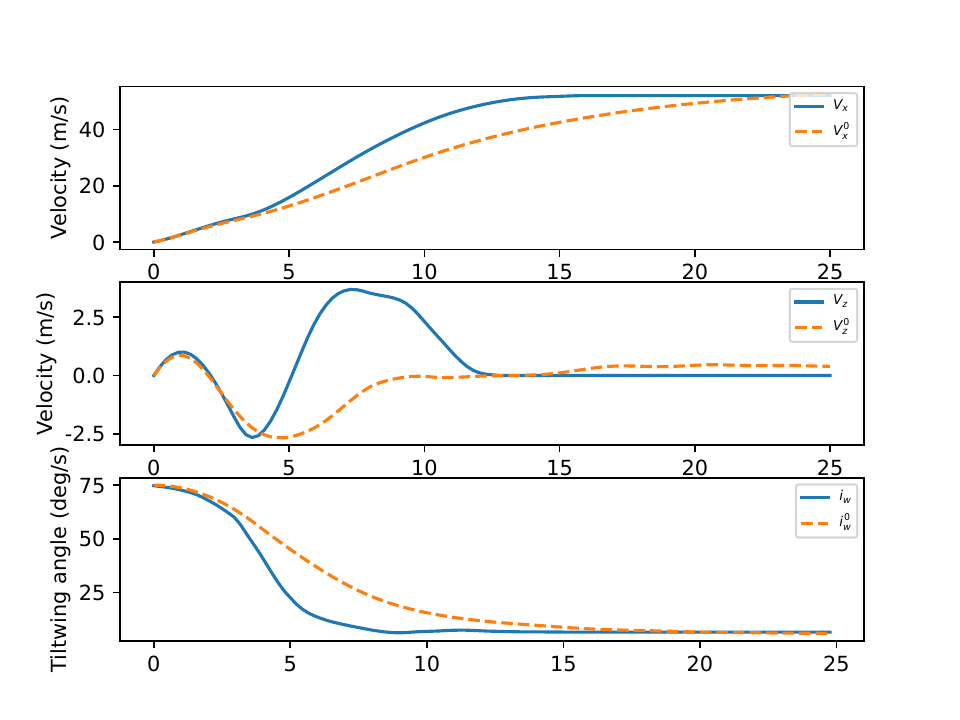}
         \caption{$t=0 s$}
         \label{fig:head0}
     \end{subfigure}
     \hfill
     \begin{subfigure}[b]{0.33\textwidth}
         \centering
         \includegraphics[width=\textwidth]{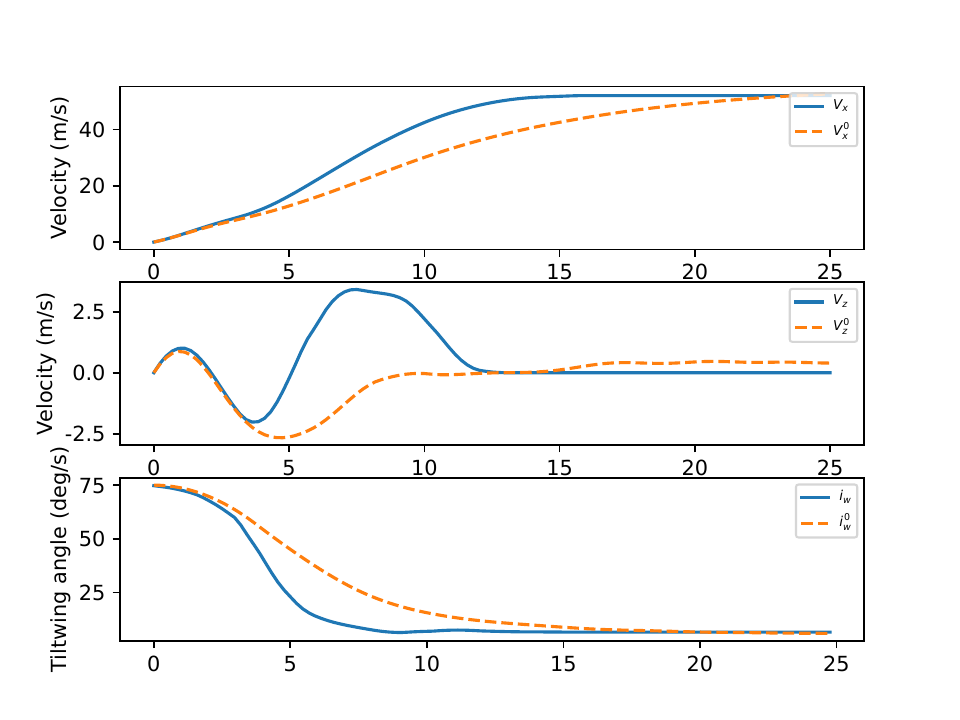}
         \caption{$t=5 s$}
         \label{fig:head1}
     \end{subfigure}
    \hfill
     \begin{subfigure}[b]{0.33\textwidth}
         \centering
         \includegraphics[width=\textwidth]{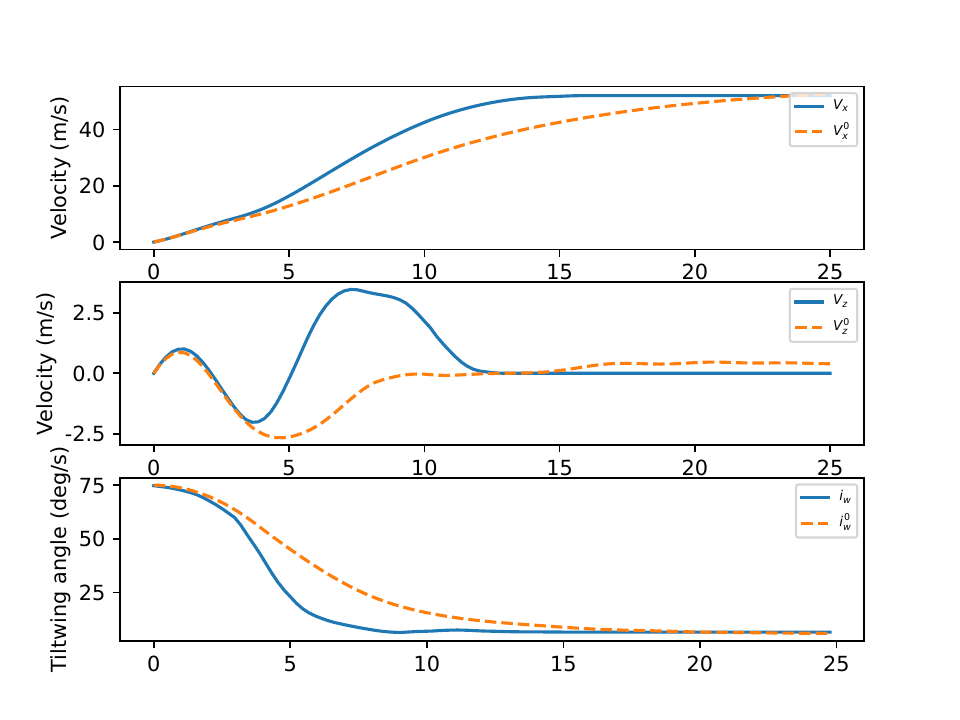}
         \caption{$t=10 s$}
         \label{fig:head2}
     \end{subfigure}
        \caption{Forward transition with headwind gust occurring at various times.}
        \label{fig:head}
\end{figure}

\begin{figure}[ht]
    \centering
    \includegraphics[width=0.6\textwidth]{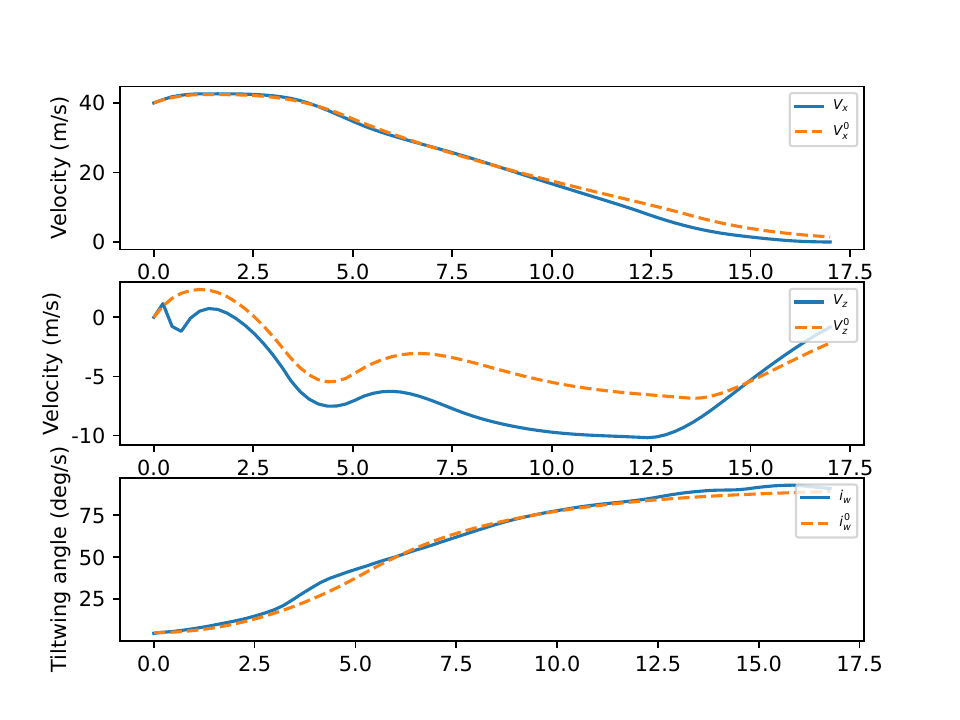} %{<left> <lower> <right> <upper>}
    \caption{Backward transition with headwind gust.}
    \label{fig:head_back}
\end{figure}

We then conduct a series of simulations of both forward and backward transitions subject to headwind gusts occurring at various time instants:

\begin{itemize}
    \item \textbf{Forward transition with headwind.} The results are presented in Figure \ref{fig:head}. As illustrated, there is almost no variation depending on when the gust is applied on the aircraft, which seems to indicate that headwinds are much less harmful than crosswinds for the closed-loop stability. This is due to the angle of attack not being affected by headwinds. 
    \item \textbf{Backward transition with headwind.} The results are presented in Figure \ref{fig:head_back}. Note that contrary to the simulations with crosswinds, the aircraft is capable to withstand headwinds at $9.14m/s$. 
\end{itemize}

\begin{comment}
\begin{figure}
     \centering
     \begin{subfigure}[b]{0.33\textwidth}
         \centering
         \includegraphics[width=\textwidth]{img/head0.pdf}
         \caption{$t=0 s$}
         \label{fig:head_back0}
     \end{subfigure}
     \hfill
     \begin{subfigure}[b]{0.33\textwidth}
         \centering
         \includegraphics[width=\textwidth]{img/head1.pdf}
         \caption{$t=5 s$}
         \label{fig:head_back1}
     \end{subfigure}
    \hfill
     \begin{subfigure}[b]{0.33\textwidth}
         \centering
         \includegraphics[width=\textwidth]{img/head2.pdf}
         \caption{$t=10 s$}
         \label{fig:head_back2}
     \end{subfigure}
        \caption{Backward transition with headwind gust occurring at various times.}
        \label{fig:head_back}
\end{figure}
\end{comment}

\begin{comment}
\begin{figure}[ht]
    \centering
    \includegraphics[width=0.6\textwidth]{img/gust0.png} %{<left> <lower> <right> <upper>}
    \caption{Forward transition with wind gust at $t=0$ $\si{s}$.}
    \label{fig:gust0}
\end{figure}

\begin{figure}[ht]
    \centering
    \includegraphics[width=0.6\textwidth]{img/gust1.png} %{<left> <lower> <right> <upper>}
    \caption{Forward transition with wind gust at $t=5$ $\si{s}$.}
    \label{fig:gust1}
\end{figure}

\begin{figure}[ht]
    \centering
    \includegraphics[width=0.6\textwidth]{img/gust2.png} %{<left> <lower> <right> <upper>}
    \caption{Forward transition with wind gust at $t=10$ $\si{s}$.}
    \label{fig:gust2}
\end{figure}
\end{comment}

\subsection{Convergence}
\begin{figure}[ht]
    \centering
    \includegraphics[width=0.6\textwidth]{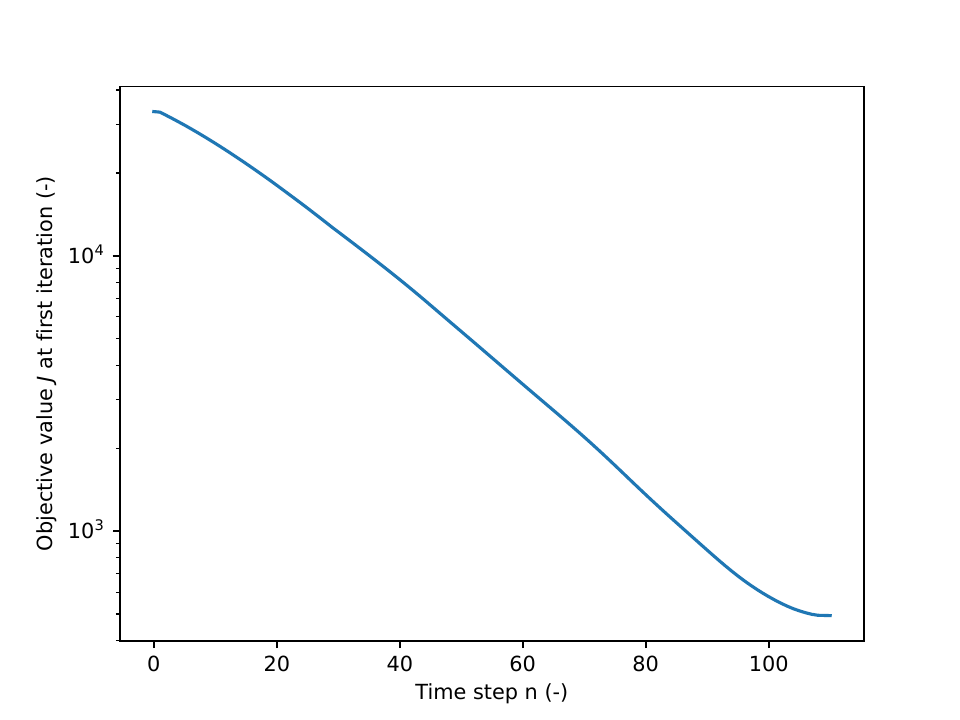} %{<left> <lower> <right> <upper>}
    \caption{Objective at each time step.}
    \label{fig:obj}
\end{figure}

\begin{figure}[ht]
    \centering
    \includegraphics[width=0.6\textwidth]{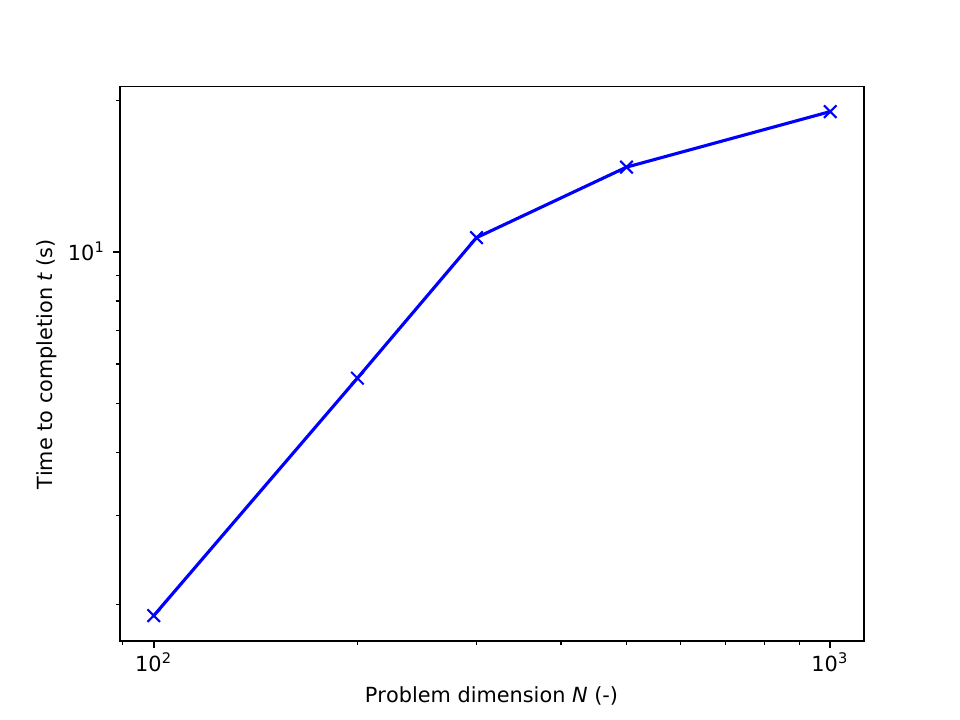} %{<left> <lower> <right> <upper>}
    \caption{Time to completion as a function of problem dimension.}
    \label{fig:time}
\end{figure}

Convergence of the algorithm is illustrated in Figure \ref{fig:obj}, showing that the objective value decreases at each time step. Finally, Figure \ref{fig:time} shows the average computation time to solve problem $\mathcal{P}$ as a function of the number of discretisation points $N$. The experiment was conducted on a MacBook Pro with a 2.9 GHz dual-core Intel Core i7 processor (mid-2012). For example, for $N=100$, the average computation time was $1.9s$. Although  this would not allow to compute the solution within the specified time step in real time, it should be noted that CVXPY is not optimised for performance and that reductions in computation times of about an order of magnitude can be expected with first order solvers such as ADMM \cite{me3}. This is in stark contrast to state-of-the-art generic NLP approaches that quote computation times of the order of minutes to solve similar VTOL transition optimisation problems (e.g. see \cite{UMich19}). 

\section{Conclusion}
\label{sec:conclusion}
We presented a novel computationally tractable robust data-driven tube MPC scheme based on a DC decomposition of the nonlinear dynamics of a tiltwing VTOL aircraft to achieve robust transitions in the presence of wind. The DC structure of the dynamics allowed us to express the MPC optimisation at each time step as a sequence of convex programs generated by successively linearising around guess trajectories and bounding tightly the effect of the necessarily convex linearisation errors. We demonstrated the viability of the scheme by considering a case study inspired from the Airbus Vahana $A^3$ VTOL aircraft using a mixture of data-based and mathematical models. Forward and backward transitions were successfully achieved, as well as transitions subject to wind gusts.  Future work has been identified as follows: i) leveraging first order solvers (e.g. ADMM) to accelerate computation times and enable real-time implementations; ii) complete study of the effect of: uncertainty set parameterisation, DC decomposition technique, approximation function, etc. on the performances of the algorithm; iii) adaptation of the method to function approximation via deep neural network  to allow a higher degree of generalisability ; iv) extension of the framework to constraints of stochastic nature (e.g. von Kármán wind turbulence model could be leveraged to achieve VTOL transitions that are less conservative). 

%\bibliography{biblio} 
%\bibliographystyle{ieeetr}
\bibliography{sample}

\end{document}